\documentclass[iop]{emulateapj}

\usepackage{ xspace, color, afterpage, natbib}
\newcommand{\kms}{ km s$^{-1}$\xspace}

\newcommand{\za}{$z_{\rm abs}$}
\newcommand{\ze}{$z_{\rm em}$}
\newcommand{\lala}{$\lambda\lambda$\xspace}
\newcommand\Lya{Lyman $\alpha$\xspace}
\newcommand\Lyb{Lyman $\beta$\xspace}

\newcommand\Lye{Lyman $\epsilon$\xspace}

\newcommand{\lstar}{L$^{\star}$\xspace}
\newcommand{\dndz}{d$\mathcal{N}$/d$z$}

\shorttitle{PG1148+549 Ne~VIII Absorbers}
\shortauthors{Meiring et al.}

\begin{document}

\title{QSO Absorption Systems Detected in Ne VIII: High-Metallicity Clouds with a Large Effective Cross Section\altaffilmark{1}}

\author{J.D. Meiring\altaffilmark{2},
T. M. Tripp\altaffilmark{2}, 
J. K. Werk\altaffilmark{3}, 
J.C. Howk \altaffilmark{4},
E.B. Jenkins\altaffilmark{5}, 
J. X. Prochaska\altaffilmark{3}, 
N. Lehner\altaffilmark{6},
K. R. Sembach\altaffilmark{6}
}

\altaffiltext{1}{Based on observations made with the NASA/ESA Hubble Space Telescope, obtained at the Space Telescope Science Institute, which is operated by the Association of Universities for Research in Astronomy, Inc., under NASA contract NAS 5-26555. These observations are associated with program GO11741.}
\altaffiltext{2}{Department of Astronomy, University of Massachusetts, Amherst, MA 01003, USA} 
\altaffiltext{3}{University of California Observatories-Lick Observatory, UC Santa Cruz, CA 95064, USA}
\altaffiltext{4}{Department of Physics, University of Notre Dame, 225 Nieuwland Science Hall, Notre Dame, IN 46556, USA}
\altaffiltext{5}{Princeton University Observatory, Peyton Hall, Ivy Lane, Princeton, NJ 08544, USA}
\altaffiltext{6}{Space Telescope Science Institute, 3700 San Martin Drive, Baltimore, MD 21218, USA}

\begin{abstract}

Using high resolution, high signal-to-noise ultraviolet spectra of the
\ze=0.9754 quasar PG1148+549 obtained with the Cosmic Origins
Spectrograph (COS) on the \textit{Hubble Space Telescope}, we study
the physical conditions and abundances of Ne~VIII+O~VI absorption line
systems at \za=0.68381, 0.70152, 0.72478. In addition to Ne~VIII and
O~VI, absorption lines from multiple ionization stages of oxygen
(O~II, O~III, O~IV) are detected and are well-aligned with the more
highly ionized species.  We show that these absorbers are multiphase
systems including hot gas ($T \approx 10^{5.7}$ K) that produces
Ne~VIII and O~VI, and the gas metallicity of the cool phase ranges
from $Z = 0.3 Z_{\odot}$ to supersolar. The cool ($\approx 10^{4}$ K)
phases have densities n$_{H}\approx 10^{-4}$ cm$^{-3}$ and small
sizes ($<$ 4 kpc); these cool clouds are likely to expand and dissipate, and the
Ne~VIII may be within a transition layer between the cool gas and a
surrounding, much hotter medium.  The Ne~VIII redshift density,
d$\mathcal{N}$/d$z\sim7^{+7}_{-3}$, requires a large number of these
clouds for every $L > 0.1L*$ galaxy and a large effective absorption
cross section ($\gtrsim 100$ kpc), and indeed, we find a star forming
$\sim$\lstar galaxy at the redshift of the \za=0.72478 system, at an
impact parameter of 217 kpc.  Multiphase absorbers like these Ne~VIII
systems are likely to be an important reservoir of baryons and metals
in the circumgalactic media of galaxies.

\end{abstract}


\section{Introduction}
The cosmic baryon fraction is now well constrained to a value
$f_b=0.17\pm0.01$ by both the interpretation of deuterium abundances
in the framework of big bang nucleosynthesis \citep{Omear06,Pettini12}
and observations of the cosmic microwave background \citep{Sper07}.
In contrast, the contributions of readily observed stellar and gas
components in nearby galaxies are far below this value
\citep{Per92,Fuk98,Bell03}.  One explanation for the ``missing
baryons'' is that substantial quantities of low density gas reside in
the intergalactic medium (IGM) or in the halos of galaxies and groups
(now generally referred to as the ``circumgalactic'' medium, CGM), and
this matter is heated during infall into the dark matter potential
wells that surround the visible galaxies \citep{Dave01, Cen06, Gar11,
  Smith11}. This shock heating is a universal prediction of
hydrodynamical simulations of structure formation, and it is expected
to transform the gas from cool ($\approx 10^{4}$ K) clouds that are
straightforward to detect \citep{Rauch97,Weinberg97} into hotter
phases ($10^{5} - 10^{7}$ K) that are much more difficult to study.
How the intergalactic gas is physically processed as it descends into
galaxies has important implications for galaxy evolution
\citep{Keres05,Keres09,Dek06,Bouche10}.  For example, it is possible
that the some of the baryons cannot cool and fall into the
star-forming disk \citep{Mall04} or are somehow prevented from
entering galactic potential wells in the first place \citep{And10}. In
any case, the observations indicate that the majority of the baryons
in the Universe are not located in the disks of galaxies; rather, they
are sequestered in highly ionized, low-density gas in galaxy halos and
the intergalactic medium (IGM). Currently, the most sensitive method
to study such low-density plasmas is to search for the ultraviolet
(UV) absorption lines imprinted on the spectrum of a background
quasistellar object (QSO) from foreground, low-density material.

In addition to probing the missing baryons, QSO absorption
spectroscopy of the CGM of galaxies provides unique insights on the
roles played by gas inflows and outflows in galaxy evolution. Simple
closed-box models of galactic chemical evolution fail to reproduce the
distribution of stellar metallicities in the Milky Way and nearby
galaxies \citep[e.g.,][]{vdb62, Lar72, Tin75, Pag81, Tosi88, Wor96}.
This well known ``G-dwarf problem'' indicates that galaxies continue
to acquire gas from their intergalactic surroundings over much of
their lifetimes.  Exactly how this gas accretion occurs is an open
question with important implications. In contrast to the traditional
picture of hot accretion through a spherical accretion shock
\citep[e.g.,][]{White78}, more recent theoretical studies suggest that
gas accretes in filamentary structures, and it may be able to cool as
it accretes so that it never approaches the virial temperature
\citep{Keres05,Keres09,Dek06,Brooks09,Fuma11,Stew11}.  Signatures of
this ``cold'' accretion are difficult to identify, but recent QSO
absorption-line studies have provided some observational evidence of
cold accretion
\citep{Tripp05,Rib11,Gia11,Thom11,Church12,Kac12,Lehner13}. These
investigations have revealed very low metallicity gas in the halos of
high metallicity galaxies; this low metallicity material is generally
cool and could arise in the accretion flows, which are expected to be
metal-poor.

Similarly, simulations and analytical models of galaxy formation and
evolution require massive \textit{outflows} to account for the
observed properties of galaxies such as the ISM metallicities and the
mass-metallicity relation, total stellar masses, discrepancies in the
galaxy luminosity function compared to expectations from cold-dark
matter cosmology, and IGM enrichment
\citep[e.g.,][]{Dek86,Spr03,Opp06}. Feedback processes from current or
recent star formation, either through radiatively driven or
supernova-driven galactic-scale winds \citep[or the combined effects
  of both processes, see][]{Mur11} are thought to play crucial roles
in the evolution of a galaxy by injecting energy and mass into the CGM
and beyond into the IGM. Such outflows are seen in nearby star forming
galaxies \citep{Heck95, Mar02, Veil05} and are ubiquitous in some
types of higher-redshift galaxies \citep{Tre07,Wein09,Ste10}, but most
of these studies provide little or no information on the full spatial
extent, total mass, and overall impact of the outflows
\citep[cf.,][]{Tripp11}, either because the source of the outflow is
used as the continuum source (and thus the spatial extent of the flow
is unconstrained) or because the studies have inadequate leverage on
the ionization and metallicity of the gas.

Which QSO absorption lines provide constraints on the missing baryons
and inflows/outflows? The O~VI \lala 1031.9, 1037.6 lines arising from
O$^{5+}$, which has a peak ionization fraction at $\sim$300,000 K in
collisional ionization equilibrium \citep{Gnat07}, have been used to
trace hot gas in a range of environments from the local ISM to the
CGM/IGM \citep[e.g.,][and references
  therein]{Jenkins78,Tripp00,Chen00,Howk02,Sembach03,Sav06,Stocke06,Bowen08,Wakker09,
  Chen09,Howk09,Leh09,Nara12}. Recent observations indicate that the extended halos
of star forming galaxies are filled with highly ionized and
metal-enriched gas traced by O~VI, and a large fraction of
star-forming galaxies have detectable O~VI absorption in their halos
out to impact parameter $\rho \approx$ 150 kpc
\citep{Tum11b}. Absorption by O~VI is also detected in the halos of
early-type galaxies, but less frequently \citep{Tum11b}, while strong
H~I absorption is ubiquitious in galaxy halos at $\rho \lesssim 150$
kpc regardless of galaxy type \citep{Thom12}. This circumgalactic
material appears to contain a substantial amount of mass, comparable
to the mass of the ISM in the galaxy itself \citep{Tum11b, Tripp11,
  Pro11}.

However, the physical nature of the O~VI-bearing gas (e.g.,
collisionally ionized vs. photoionized) is a debated and open question
\citep{Tripp08}, a question with important
ramifications. Hydrodynamical simulations of galaxy evolution show
that O~VI absorbers typically reside in metal enriched regions with
overdensities $\rho/\langle\rho\rangle = $ 1$-$100 \citep{Cen01,Fang01,
  Cen06, Dave01, Gar11, Opp09, Cen11,Opp11, Smith11}, and these models have
made detailed predictions regarding the strength and physical nature
of O~VI absorbers as a function of impact parameter and galaxy
luminosity \citep{Gang08, Cen11, Stinson12, Ford12}. Simulations by
\citet{Gar11} indicate that O~VI does appear to trace primarily shock
heated gas, with the distribution of temperatures peaked at
$\sim10^{5.3}$ K.  However, $\sim$30 percent of the O VI systems in
that simulation are at lower temperatures and are \textit{primarily}
photoionized by the ultraviolet background flux. Other simulations
indicate that O~VI and even Ne~VIII absorbers\footnote{We note,
  however, that most of the Ne~VIII absorbers predicted by
  \citet{Opp11} have Ne~VIII column densities that are much lower than
  the observed Ne~VIII column densities that have been reported thus
  far; the stronger Ne~VIII systems that have been observed are likely
  to originate in collisionally ionized gas in the simulations of
  \citet{Opp11}.}  arise primarily in photoionized gas \citep{Opp09,
  Opp11}.  Interestingly, $\sim30\%$ of the \textsl{observed} O~VI
absorbers have characteristics of cool, photoionized gas
\citep{Tripp08}.

Despite recent advances in the UV absorption spectroscopy of the
low-$z$ IGM, several questions are still unanswered, due in part to
the lack of adequate diagnostic lines at low redshift.  The baryonic
content, metallicity, and ionization mechanisms of the gas comprising
the IGM and CGM are only loosely constrained, and the roles played by
circumgalactic and intergalactic gases in galaxy evolution are still
highly uncertain.  Much of the work on low-$z$ absorption lines has
been accomplished with the \textit{Hubble Space Telescope} Space
Telescope Imaging Spectrograph (STIS) and \textit{Far Ultraviolet
  Spectroscopic Explorer} (FUSE); these spectrographs can only study
the brightest QSOs in reasonable exposure times. At the redshifts of
the absorbers that have typically been studied with STIS and
\textit{FUSE}, only a limited set of absorption lines is available for
analysis, and much of the determination of the physical conditions in
the systems is based on H~I and O~VI alone, or on column density
ratios of different ions such as C~III, Si~III, and
O~VI. Consequently, there can be degeneracies between effects due to
ionization, non-solar relative abundances, and depletion by dust.
Moreover, with a limited set of transitions, key species in the
analysis (e.g., C~III and Si~III, which have only one strong resonance
transition at $\lambda > 912$ \AA ) are prone to saturation.

In order to overcome some of these issues and to better constrain the
physical conditions in the IGM and CGM, we have undertaken a blind
survey of $z\sim1$ QSOs with the COS spectrograph \citep{Fro09,
  Green12}. At moderate redshifts, we can access many of the numerous
absorption lines in the FUV at $\lambda_{\rm rest} < 912$
\AA\ \citep{Verner94}, which provides several advantages compared to
previous studies: First, multiple and adjacent ionization states of
oxygen as well as many other ions (e.g., O~I -- O~VI, N~I -- N~V,
S~II -- S~VI, etc.) are available at $\lambda \gtrsim$ 600 \AA\ in the
rest frame so that diagnostics of physical conditions ranging from $T
\approx 10^{3}$ K up to $T \gg 10^{5}$ K can be exploited without
confusion from assumptions about the relative abundances of the
elements.  Likewise, relative abundances can be constrained with the
ambiguity from ionization corrections greatly reduced.  Finally, in
this wavelength range, many species have multiple transitions so that
if a strong line is saturated or a line is badly affected by blending,
other transitions can still be used to measure the column density of
interest.  At $z\sim0.5$, these transitions enter the bandpass of the
Cosmic Origins Spectrograph (COS) onboard the \textit{Hubble Space
  Telescope} (HST).

Aside from O~VI, other highly ionized species have resonance
transitions that are accessible in the FUV. For example, the Ne VIII
\lala 770,780 doublet enters the COS bandpass at $z\gtrsim 0.45$.  In
diffuse halo gas or the intergalactic medium, it is difficult to
photoionize neon up to Ne~VIII: the ionization potential of Ne~VII is
207 eV and the ionization potential of Ne~VIII is 239 eV. Neither
stars nor background quasars produce an ionizing flux field with many
photons at these energies \citep[cf.][]{Fox05,Haardt12}, and
photoionization models typically require very low densities and, in
turn, very large clouds to yield detectable quantities of Ne~VIII
\citep{Sav05,Nar11}.  While Ne~VIII is difficult to produce by
photoionization, it can easily originate via collisional ionization in
gas that is sufficiently hot. Consequently, Ne~VIII is expected to be
an unambiguous indicator of collisionally ionized hot gas.  Several
intervening\footnote{The Ne~VIII doublet has also been detected in a
  number of ``proximate'' absorbers with $z_{\rm abs} \approx z_{\rm
    QSO}$ \citep[e.g.,][]{Petit99,Gang06, Muza12, Muza13}; these proximate
  absorbers show evidence that they are close to the central engine of
  the QSO and thus probe a different aspect of galaxies than the
  intervening systems. In this paper, we are primarily interested in
  diffuse gas in the halos of galaxies and the IGM, so we do not
  include the proximate systems in our discussions.} absorption
systems bearing Ne VIII have been discovered, indicating highly
ionized and multiphase gas with temperatures of T=10$^{5} - 10^{6}$ K
in the hot material \citep{Sav05,Tripp11,Nar11,Nara12}.

The excellent sensitivity of COS in the UV provides an opportunity to
study the IGM and CGM in much more detail than has been achievable in
the past.  Using our high signal-to-noise COS spectrum of the QSO
PG1148+549, in this paper we study in detail absorption systems at
$z\sim0.7$ where we have access to a number of ionization states of
oxygen (O I to O VI) and the Ne VIII \lala 770, 780 doublet; we also
place limits on other banks of (undetected) adjacent ions such as the
sulfur ions.  We will show that the O~VI and Ne~VIII does indeed arise
in plasma with $T \gg 10^{5}$ K, but this hot gas has an intimate
relationship with lower ionization material -- the hot gas likely
originates in some sort of interface on the surface of the cooler,
low-ionization clouds \citep[e.g.,][]{Kwak11}.  We will also show that
these Ne~VIII systems are remarkably metal-rich, and when information
on nearby galaxies is available, the Ne~VIII systems tend to be
surprisingly far from luminous galaxies.  The organization of the
paper is as follows: In $\S$ 2, we discuss the COS observations and
data handling, details of the analysis of each absorption system are
given in $\S$ 3, $\S$ 4 describes the ionization modeling, results are
given in $\S$ 5, a discussion of the galaxies in the field are given
in $\S$ 6, and $\S$ 7 gives a summary and discussion.  Throughout this
paper we assume the 737 cosmology: $H_{0}$ = 70 km s$^{-1}$
Mpc$^{-1}$, $\Omega _{m}$ = 0.3, and $\Omega _{\Lambda}$ = 0.7.

\vspace{1cm}
\section{Observations}
\label{Sec:Observations}

\begin{table}
\begin{center}
\caption{ Cosmic Origins Spectrograph Observations of PG1148+549 \label{Tab:COS_Obs}}
\begin{tabular}{lcccc}
\hline \hline
COS Grating   & $\lambda_{cen}$    & Observation & Exposure   & MAST ID$^{\rm a}$ \\
   \          &                 & Date           & Time       &           \\
   \          & (\AA )     & \              & (seconds)  &           \\ \hline
G130M\dotfill    & 1309       & 2009 Dec. 30   & 3192       & LB1O22Q3Q \\
G130M\dotfill    & 1309       & 2009 Dec. 30   & 3192       & LB1O22RAQ \\
G130M\dotfill    & 1309       & 2009 Dec. 30   & 2527       & LB1O22PMQ \\
G130M\dotfill    & 1327       & 2009 Dec. 26   & 2527       & LB1O23EBQ \\
G130M\dotfill    & 1327       & 2009 Dec. 30   & 3192       & LB1O22RHQ \\
G130M\dotfill    & 1327       & 2009 Dec. 30   & 3192       & LB1O22RRQ \\
G160M\dotfill    & 1600       & 2009 Dec. 26   & 3192       & LB1O23EOQ \\
G160M\dotfill    & 1600       & 2009 Dec. 26   & 3192       & LB1O23EFQ \\
G160M\dotfill    & 1600       & 2009 Dec. 26   & 3192       & LB1O23EVQ \\
G160M\dotfill    & 1623       & 2009 Dec. 25   & 2482       & LB1O24ACQ \\
G160M\dotfill    & 1623       & 2009 Dec. 25   & 3192       & LB1O24AFQ \\
G160M\dotfill    & 1623       & 2009 Dec. 26   & 3192       & LB1O23F2Q \\ 
\hline
\end{tabular}
\end{center}
$^{\rm a}$ Exposure identification code in the Mikulski Archive for Space Telescopes (see http://archive.stsci.edu/index.html).
\end{table}

The observations of PG1148+549 presented here were taken as part of
\textit{HST} program 11741 (PI Tripp), a blind survey for highly
ionized species such as O~VI, Ne~VIII, and Mg~X, as well as lower
ionization stages that constrain the physical conditions and
metallicity of the multiphase galactic halos, in the spectra of $z\sim
1 - 1.5$ UV-bright quasars. Targets for this sample were selected only
based on their redshift ($z \ge 0.9$) and flux in the far ultraviolet
(FUV). No consideration was given to previously known information
about absorption systems when selecting targets for this program, with
two exceptions: (1) If a target was known to have a broad absorption
line system in its spectrum, it was excluded because such systems,
which are known to be ejected gas located very close to the QSO
nucleus, have complex and temporally variable absorption profiles that
spread over large wavelength ranges \citep[see, e.g.,][]{Trump06} and
seriously compromise our ability to study \textit{intervening} gaseous
halos and intergalactic gas clouds. (2) If a target was known to have
strong Lyman limit absorber that completely suppresses the flux
shortward of its redshifted Lyman limit in some portion of the COS FUV
band, the target was excluded in order to maximize the wavelength
range that can be usefully searched for the species of
interest.\footnote{Note that a target was not rejected if its spectrum
  showed a \textit{partial} Lyman limit, i.e., only a fully black
  Lyman limit absorption led to rejection.}  While PG1148+549 had been
previously observed with the \textit{HST} Faint Object Spectrograph
(FOS, $R\sim1300$) in programs 4952 and 6210 \citep{Ham98,Bech02}, the
resolution and sensitivity of the FOS spectra are far too low for the
purposes of this work, which requires detection of $\approx 20$
m\AA\ lines and precise and reliable measurement of line centroids,
line widths, and profile kinematics/component structure.

As summarized in Table \ref{Tab:COS_Obs}, the COS spectra of
PG1148+549 were acquired on 25-30 December 2009. To cover the full FUV
wavelength range of COS from 1150 to 1800 \AA , observations were
obtained with both the G130M and G160M gratings, and for each grating
two central wavelength tilts were used to fill in the gap between the
two COS detector segments. In addition, several exposures with
multiple focal-plane positions (FP-splits) were obtained with each
central-wavelength setting so that the effects of fixed-pattern noise
are mitigated when the individual exposures are aligned and combined.
Overall, the total exposure times were estimated based on the goal of
detecting Ne~VIII lines as weak as those reported by \citet{Sav05},
i.e., exposure times were calculated to achieve adequate
signal-to-noise (S/N) to detect lines with equivalent widths of 20 --
30 m\AA\ with good statistical significance over the full wavelength
region where the doublet is redshifted into the \textit{HST}
bandpass.

The COS data were processed in the same manner as described in
\citet{Mei11}. The COS FUV detector backgrounds are very low
\citep{Green12}, so in the cores of strong absorption lines, the total
counts can be low enough so that Poissonian statistics should be used
to estimate the flux uncertainties.  Since COS has photon-counting
detectors, we used the counts in each pixel to determine flux
uncertainties, using Poissonian statistics in regions of low counts.
In addition to multiple FP-split exposures, flatfielding was applied
to further reduce residual fixed-pattern noise, primarily from the COS
gridwires.  Strong and narrow absorption lines were used to cross
correlate and align the individual exposures, and the reduced and
coadded spectra were binned by 3 pixels since the standard pipeline
COS data are oversampled with a $\sim6$ pixel wide resolution element
(i.e., the data were binned to Nyquist sampling of $\sim $2 pixels per
resolution element). All measurements and analyses in this paper were
performed on the binned spectra. The binned spectra have a resolution
of $15 - 20$ \kms per resolution element, and the signal-to-noise
ratio of the final, fully combined COS spectrum ranges from 20 to 40
per resolution element, with S/N $>$ 29 in most continuum pixels
between 1180 and 1550 \AA . 

In principle, there are several potential sources of noise in addition
to photon-counting statistics such as uncertainties in the continuum
placement, uncertainties in the flux zero point, and fixed-pattern
noise.  Due to the low detector backgrounds of COS, uncertainties in
the flux zero point are very small and can be safely neglected.  We
also expect the fixed-pattern noise contribution to be small due to
our use of FP splits and flatfielding. To check this, we have measured
the RMS noise in line-free continuum regions across the full
wavelength range of the spectrum, and we find that the noise in the
continuum is in good agreement with the expected noise based on
photon-counting statistics, which indicates that fixed-pattern noise
makes a small and generally unimportant contribution to the overall
measurement uncertainties.  However, in some regions,
continuum-placement uncertainty can be comparable to the statistical
noise, so we have included this term in our error analyses.

While the COS FUV absorption lines are the central data of this
survey, we also obtained several ancillary observations from the
ground to support the COS analyses:  

First, to search for Mg~II $\lambda \lambda 2796,2803$ absorption
affiliated with systems of interest, we obtained high-resolution
optical spectra of PG1148+549 with the High Resolution Echelle
Spectrograph (HIRES) at the Keck Observatory.  Two 900-second
exposures were recorded with HIRESb on 2012 April 13 covering the 3300 $-$ 5880 \AA\ range with a spectral
resolution of 6 km s$^{-1}$ (FWHM).  In the wavelength range relevant
to the absorption systems studied in this paper, the HIRES data have
S/N = 40 to 50 per resolution element at the expected wavelengths of
the Mg~II doublet. 

Second, we obtained deep multiband imaging of the PG1148+549 field
with the twin Large Binocular Cameras (LBCs) on the 2$\times$8.4-m
Large Binocular Telescope (LBT).  The LBCs, fully described in
Giallongo et al. (2008), provide simultaneous imaging in two bands
over a $\sim23\arcmin$ field of view centered on the QSO.  We obtained
imaging in the UBVI bands, dithering between exposures to fill in the
inter-chip gaps between the four CCDs.  The total U and I band
exposures total 2500 sec, while the B and V band exposures total 350
sec.  The imaging was taken through very light cirrus with $\sim1.0''$
seeing.  In the current paper we make use of the LBC blue-side
imaging; the U and B-band images reach 5$\sigma$ limiting magnitudes
of $U_{AB}\sim26.0$ and $B_{AB}\sim25.5$ mag in a $2\arcsec$ aperture.
This is roughly equivalent to $L \sim 0.1 - 0.2 \, L*$ at $z\sim0.7$.

Finally, using the instrument setup and data-reduction procedures of
\citet{Werk12}, we measured the redshifts, star-formation rates, and
metallicities of galaxies close to the PG1148+549 sight line using the
Keck Low Resolution Imaging Spectrometer (LRIS).  Four of the
brightest objects within 45'' of the sight line were targeted with LRIS,
with exposure times ranging from 600 to 800 s, on 2010 April 5.  The
LRIS spectra verified that two of the targets are stars, but the other
two objects are galaxies, including a galaxy at the redshift of one
the Ne~VIII absorbers.  

\section{Absorption-Line Measurements}
\label{sec:ab_line_meas}

After normalizing the continuum in regions of interest by
interactively fitting cubic splines to the data, we employed two
techniques for absorption-line measurements.  First, we used the
apparent column density method \citep{Sav91,Jenk96} in which the
apparent optical depth in a pixel, $\tau _{\rm a}(v) =
\textrm{ln}[I_{\rm c}/I(v)],$ where $I(v)$ is the observed flux and
$I_{\rm c}(v)$ is the continuum flux at velocity $v$, is used to
estimate the column density in that pixel, $N_{\rm a}(v) = (m_{\rm
  e}c/\pi e^{2}) \tau _{\rm a}(v)/(f\lambda)$, where $f$ is the
oscillator strength and $\lambda$ is the wavelength of the transition,
and the other symbols have their usual meanings.  Apparent column
density profiles are useful in a variety of ways: (1) the profiles
provide an efficient means to confirm line identifications, (2)
$N_{\rm a}(v)$ profiles can be scaled and overplotted to visually
compare the kinematics and component structure of various species, and
(3) the apparent column density profiles provide a quick assessment of
whether lines are affected by saturation (through comparison of weak
vs. strong transition of a specific species).  If a profile is not
affected by saturation, it can be integrated to obtain the total
column density, $N$(total) = $\int N_{\rm a}(v) dv$.  

Second, we used Voigt profile fitting to determine the column
densities, velocity centroids, and b-values of H~I and metal
absorption lines detected in the PG1148+549 spectrum.  To account for
the broad wings in the COS line-spread function
\citep{Ghav09,Green12}, the synthetic Voigt profiles were convolved
with the COS line spread functions (LSF)
at the nearest tabulated wavelength \citep{Ghav09}.  Whenever
possible, multiple transitions were fit simultaneously (e.g. Lyman
$\beta , \gamma , \delta$) to determine the best fitting
column densities.  

This paper is focused on Ne~VIII/O~VI absorption systems, but since
previous studies have always found that these systems also exhibit
absorption from lower ionization stages, we have also searched for
absorption lines from the strongest species expected in warm ionized
gas as well. In the COS data, we search for ions of
carbon, nitrogen, oxygen, and sulfur, and we use the Keck spectra to
place upper limits on Mg~II.  Upper limits on column densities for
non-detected species were derived by determining a $3\sigma$ upper
limit on the rest-frame equivalent width of the undetected ion,
including an allowance for continuum placement uncertainty using the
method of \citet{Sem92}, and then the equivalent width upper limit was
converted to a column density upper limit assuming the linear curve of
growth applies.  Often the COS data cover multiple transitions of a
given ion; our strategy for placing upper limits is to use the
strongest transition (and thereby obtain the tightest constraint) that
is not badly affected by blending with unrelated lines.

The COS wavelength solution is known to have inaccuracies at the level
of one to two (binned) pixels, and these inaccuracies are particularly
noticable in regions of the spectra that are recorded near the edges
of the detector segments \citep{Sav11}; the worst errors can approach
30 km s$^{-1}$, although typically the errors are smaller than this.
In some cases, we are able to correct for this wavelength calibration
problem by comparing lines that should be well-aligned (e.g., H~I
Lyman series lines or different transtions of the same metal ion), and
we have applied shifts in velocity to improve the wavelength
calibration when such comparisons have been possible.  We define the
systemic redshifts of the absorbers to be the centroid of the
strongest component of the O~VI profiles.

\section{\textsc{Ne}~VIII Absorption Systems}

\begin{table}
\caption{Equivalent Widths and Integrated Apparent Column Densities for Transitions Observed in the \za=0.68381 System. 
               \label{Tab:AOD6838}}
\begin{center}
\begin{tabular}{lcc}
\hline\hline
  Transition                    & $W_{\rm r}$ (m\AA ) & log [$N_{\rm a}$ (cm$^{-2}$)]     \\
\hline
H I $\lambda$1025.72\dotfill    & 276$\pm$17          &    14.77$\pm$0.02     \\
H I $\lambda$949.74\dotfill     & 54$\pm$15           &    14.74$\pm$0.11      \\
H I $\lambda$937.80\dotfill     & 37$\pm$9            &    14.82$\pm$0.09             \\
C~II $\lambda$903.62$^{\rm b}$\dotfill    & $<19^{\rm c}$       & $<13.2^{\rm d}$ \\
O II $\lambda$834.47\dotfill    & $<24^{\rm c}$       & $<13.5^{\rm d}$   \\
Mg~II $\lambda$2796.35\dotfill  & $<20^{\rm c,e}$     & $<11.7^{\rm d}$ \\
N~III $\lambda$685.51\dotfill   & $<34^{\rm c}$       & $<13.5^{\rm d}$ \\
C III $\lambda$977.02\dotfill   & 188$\pm$12          &     13.65$\pm$0.02              \\
S~III $\lambda$698.73\dotfill   & $<15^{\rm c}$       & $<12.7^{\rm d}$   \\
N IV $\lambda$765.15\dotfill    & 80$\pm$8            &    13.47$\pm$0.04                \\
O IV $\lambda$787.71\dotfill    & 153$\pm$8           &    14.60$\pm$0.02                \\
S~IV $\lambda$809.67\dotfill    & $<15^{\rm c}$       & $<13.4^{\rm d}$   \\
O VI $\lambda$1031.93\dotfill   & 234$\pm$19          &     14.47$\pm$0.03                 \\
O VI $\lambda$1037.62\dotfill   & 149$\pm$23          &     14.50$\pm$0.06               \\
S~VI $\lambda$944.52\dotfill    & $<35^{\rm c}$       & $<13.3^{\rm d}$ \\
Ne VIII $\lambda$770.41\dotfill & 51$\pm$12           &    13.98$\pm$0.09              \\
\hline
\end{tabular}
\end{center}

$^{\rm a}$ {\footnotesize The O~III $\lambda$702.33 and
  $\lambda$832.93 lines are detected but are not listed in this table
  because both of these transitions are significantly blended with
  unrelated absorption features (see Figure~\ref{Fig:vel6838}), and
  thus the integrated quantities are not very useful.  We estimate the
  O~III column densities by jointly fitting Voigt profiles to both
  O~III lines as well as the blended lines.} \\
$^{\rm b}$ {\footnotesize We use the C~II $\lambda$903.62 instead of the stronger C~II $\lambda$903.96 transition because the 903.96 \AA\ line is blended into the wing of a strong Ly$\alpha$ line at $z_{\rm abs}$ = 0.2525. \\
$^{\rm c}$ {\footnotesize Three $\sigma$ upper limit obtained by integration over $-75 < v < 100$ km s$^{-1}$.}} \\
$^{\rm d}$ {\footnotesize Upper limit based on the $3\sigma$ upper limit on $W_{\rm r}$, assuming the line is in the linear regime of the curve of growth.} \\
$^{\rm e}$ {\footnotesize Derived from the Keck/HIRES data.}
\end{table}

\begin{table}
\caption{Column Densities from Profile Fits to the COS Data for the \za=0.68381 System.  \label{Tab:Cols6838}}
\centering
\begin{tabular}{lrcc}
\hline\hline
  Ion           & $v_{rad}$ \ \ \ \ & log [N (cm$^{-2}$)]   & $b$      \\
                & (\kms\ )  &                       & (\kms\ )         \\
\hline
H I\dotfill     & 0  \ \ \ \ \      &   14.65 $\pm$0.02           &           39.9$\pm$2.4            \\
                & 40 \ \ \ \ \      &   14.24$\pm$0.05            &           20.0$\pm$3.2            \\
C III\dotfill   & 0  \ \ \ \ \      &   13.57$\pm$0.02            &           22.5$\pm$1.7             \\ 
                & 40 \ \ \ \ \      &   13.16 $\pm$0.07           &             8.5$\pm$2.2             \\
O III\dotfill   & 0  \ \ \ \ \      &     14.02$\pm$0.03          &           20.5$^{\rm a}$                        \\   
                & 40 \ \ \ \ \      &      13.64$\pm$0.06         &            16.2$^{\rm a}$                        \\   
O IV\dotfill    & 0  \ \ \ \ \      &      14.49$\pm$0.03         &            20.5$\pm$1.8             \\
                & 40 \ \ \ \ \      &       14.28$\pm$0.04       &            16.2$\pm$2.0             \\
O VI\dotfill    & 0  \ \ \ \ \      &     14.41$\pm$0.02        &            31.5$\pm$1.9            \\
                & 40 \ \ \ \ \      &     13.88$\pm$0.06        &           31.0$\pm$5.0              \\
N IV\dotfill    & 0  \ \ \ \ \      &      13.32$\pm$0.05        &           23.9$\pm$2.7              \\
                & 40 \ \ \ \ \      &       12.96$\pm$0.07      &           22.1$\pm$5.3               \\ 
Ne VIII\dotfill & 0  \ \ \ \ \      &     13.95$\pm$0.04        &            32.0$\pm$5.0              \\
                & 40$^{\rm b}$ \ \ \ \ \      &                  --              &                       --                            \\
\hline
\end{tabular}
\begin{minipage}{\linewidth}

$^{\rm a}$Due to the blending in this line, the $b$ values for the
  components were fixed to the values determined from the O IV
  $\lambda$ 787 line.  We report the column density uncertainties
  formally derived by the profile-fitting procedure, but we note that
  these lines may suffer from larger systematic uncertainties due to
  the significant blending with unrelated absorption.

$^{\rm b}$ The Ne~VIII $\lambda$770.41 line is weak, and while its
  velocity width and shape are consistent with those of the stronger
  metal lines, the Ne~VIII $\lambda$770.41 profile does not provide
  enough leverage, at the S/N provided by our COS spectrum, to enable a
  two-component fit (see Figure~\ref{Fig:vel6838}), so we have fitted
  the profile with a single line at $v = 0$ km s$^{-1}$.

\end{minipage}
\end{table}

\begin{figure*}
\plotone{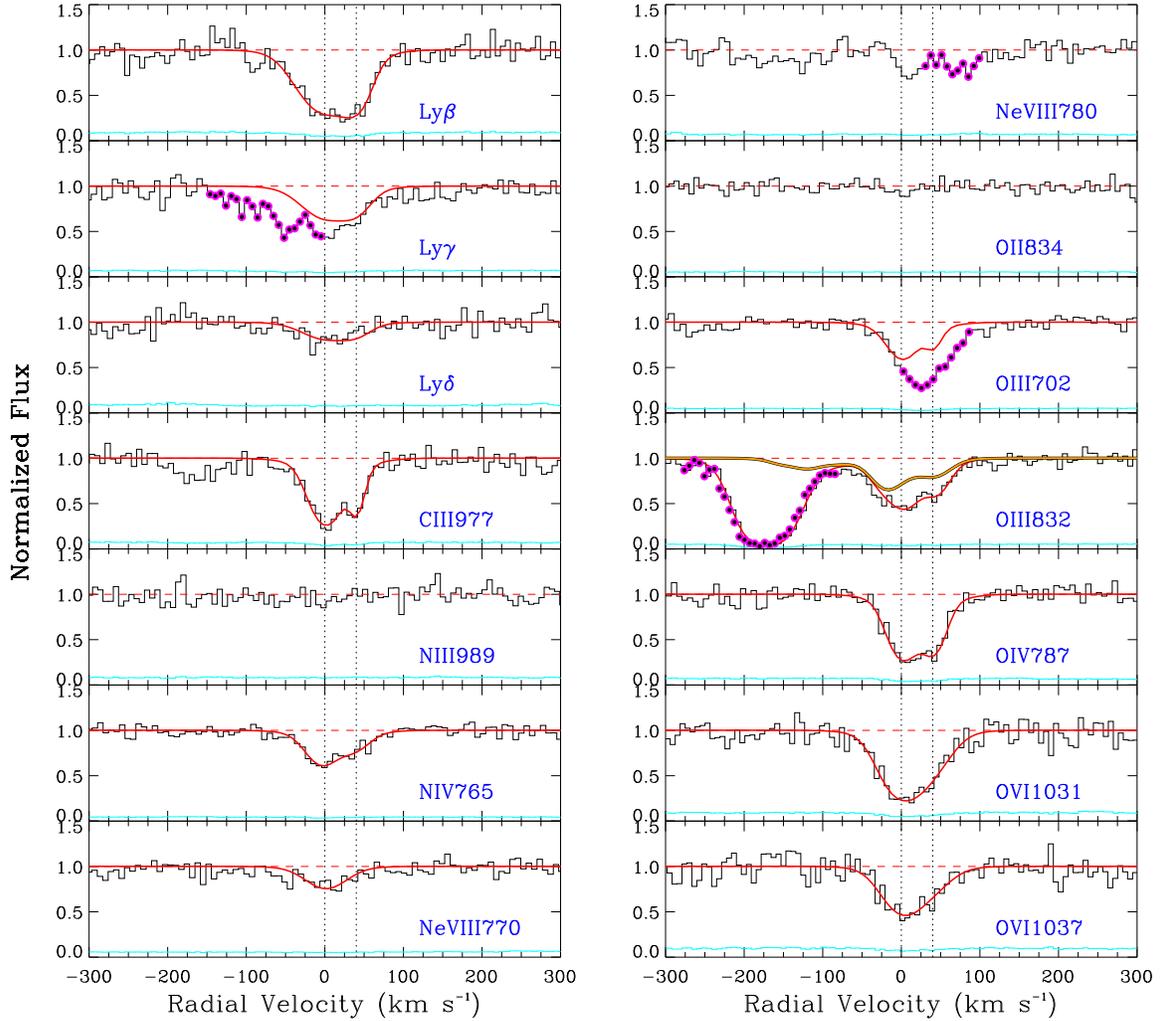}
\caption{Continuum-normalized velocity plots of the
  absorption lines in the PG1148+549 Ne~VIII absorber at \za=0.68381,
  plotted in the rest frame of the absorber (i.e., $v = 0$ km s$^{-1}$
  at \za=0.68381). Magenta points denote regions where the lines are
  blended with unaffiliated features from other systems, and the cyan
  line at the bottom of each panel denotes the 1$\sigma$ error of the
  normalized flux.  Vertical dotted lines mark the positions of the
  components, and the solid red line represents the best fit models
  determined from profile fitting.  The O III $\lambda$ 832 line is
  partially blended with the Si IV $\lambda$ 1402 line from the Milky
  Way. The contribution from Si IV $\lambda$ 1402 (based on the
  unblended Galactic Si~IV $\lambda$1393 profile) is shown with the
  solid orange line. \label{Fig:vel6838}}
\end{figure*}

Using the line-identification procedure of \citet{Tripp08}, we have
identified three Ne~VIII absorption systems in the spectrum of
PG1148+549 at $z=$0.68381, 0.70152, 0.72478.  In addition to Ne~VIII,
we detect multiple ionization stages of oxygen, nitrogen, and carbon
in these absorption systems, and we are able to place strong
constraints on the H~I content of the absorption systems, at least for
the cooler phases.  In this section we present basic information and
measurements for these Ne~VIII systems. 

\begin{figure}
\includegraphics[width=3.5in,height=2.5in]{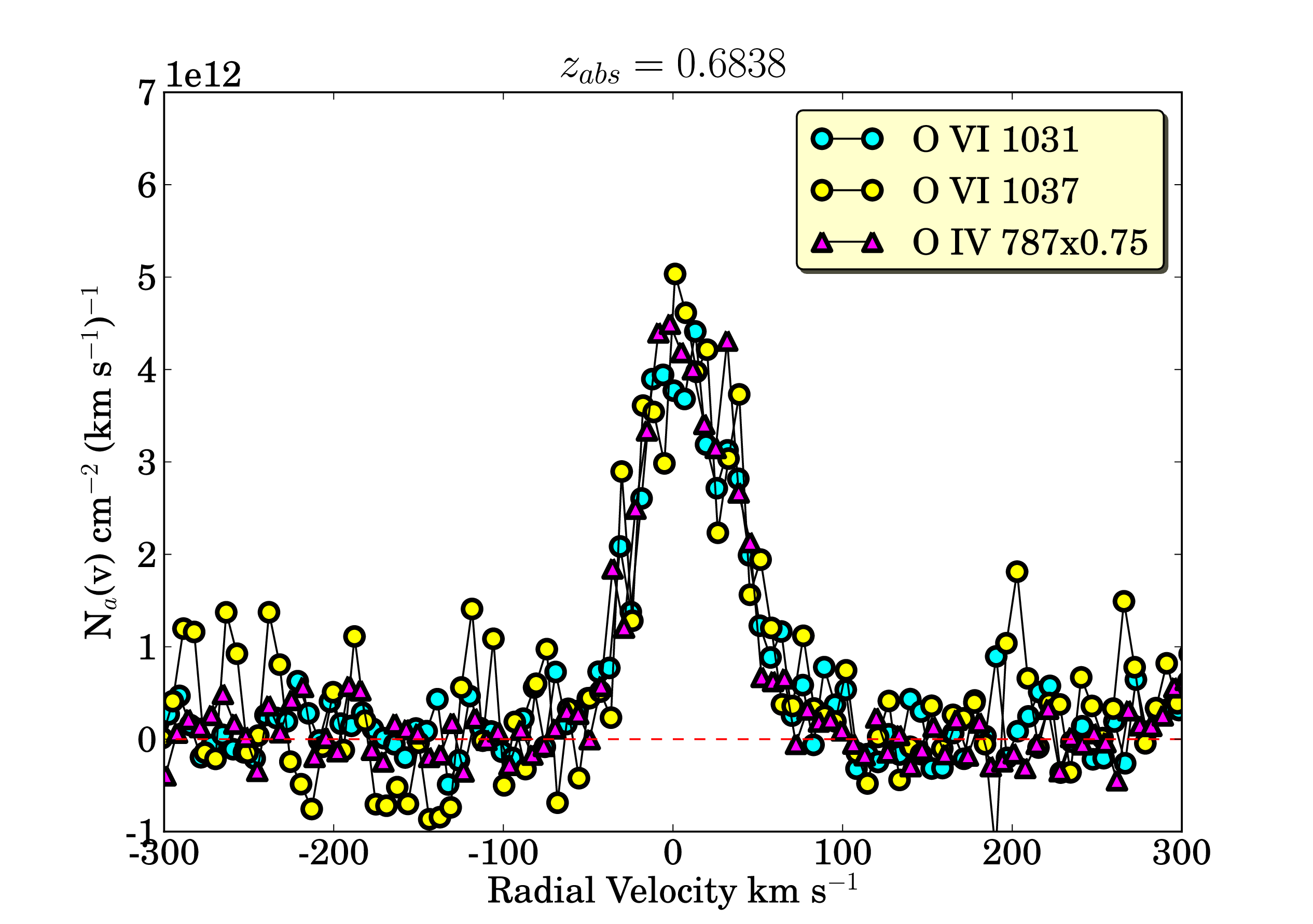} 
\caption{Apparent column density profiles (\S~\ref{sec:ab_line_meas})
  of the O~IV and O~VI lines from the \za=0.68381 system. The O~IV
  $\lambda$ 787 line has been scaled by a factor of 0.75 for
  illustration. As in Figure~\ref{Fig:vel6838} and all velocity plots
  in this paper, the velocity scale is in the rest frame of the
  absorption system. The y axis is plotted in units of $10^{12}$
  particles cm$^{-2}$ (km s$^{-1}$)$^{-1}$.\label{Fig:AOD6838_1} }
\end{figure}

\begin{figure}
\includegraphics[width=3.5in,height=2.5in]{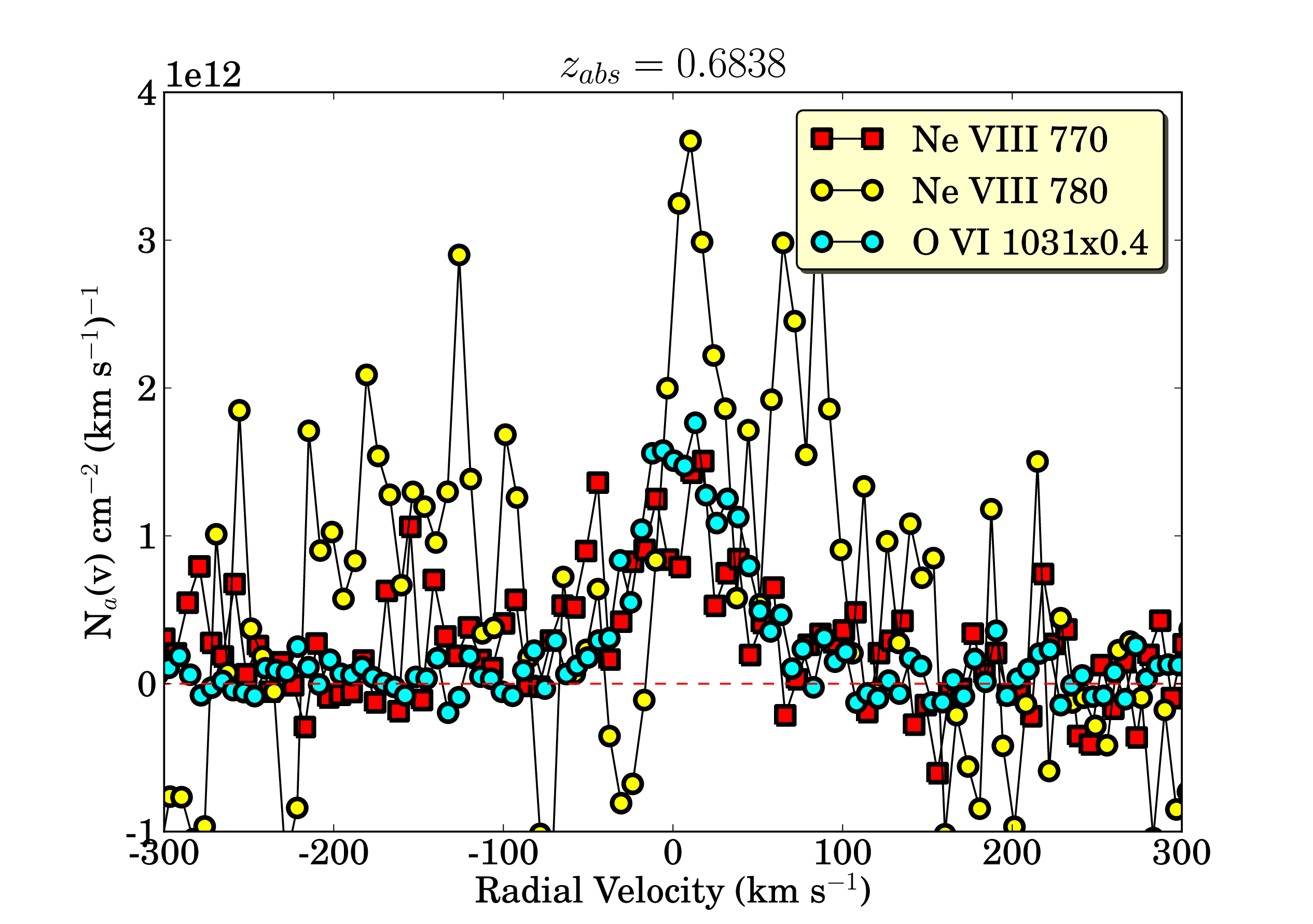}
\caption{Apparent column density profiles of the O VI and Ne VIII
 lines from the \za=0.68381 system. The O VI $\lambda$ 1031 line has
 been scaled by a factor of 0.4 for illustration. \label{Fig:AOD6838_2} }
\end{figure}

\begin{figure*}
\plotone{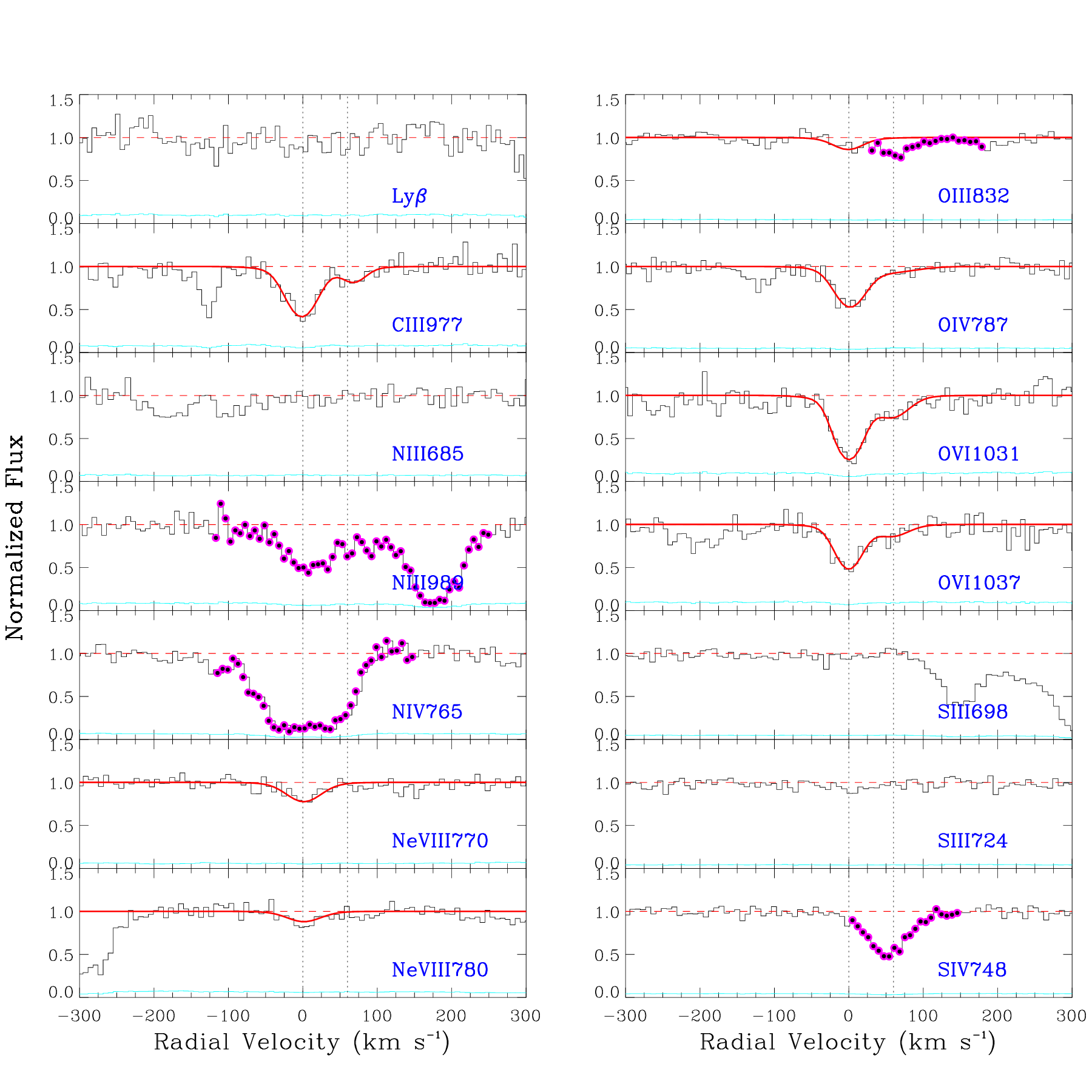}
\caption{Continuum-normalized absorption profiles of the \za=0.70152
  absorption system plotted in the rest frame of the absorber, as in
  Figure~\ref{Fig:vel6838}.  \label{Fig:vel7015}  }
\end{figure*}

\begin{figure}
\includegraphics[width=3.5in,height=2.5in]{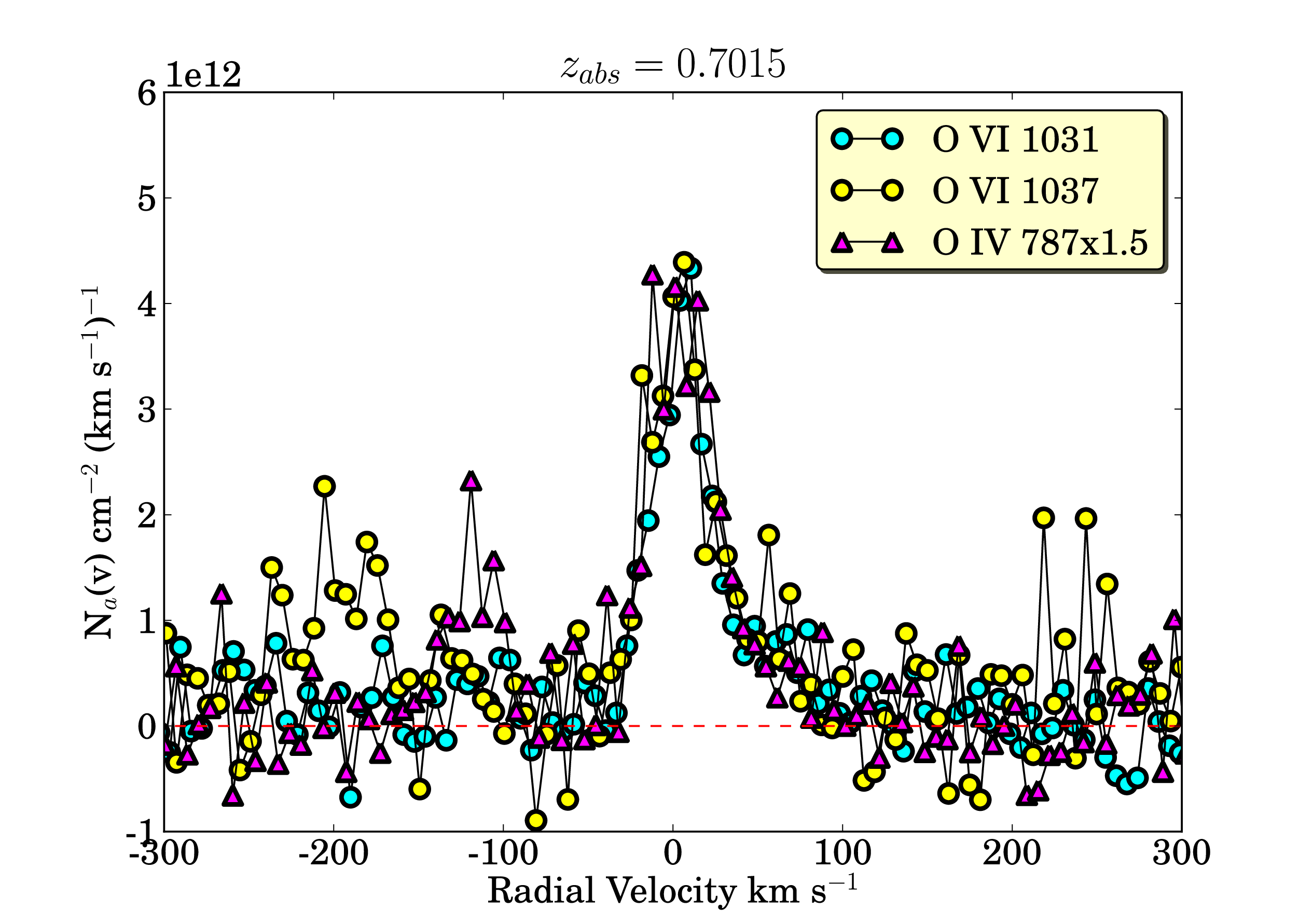} 
\caption{Apparent column density profiles of the O~IV  and O~VI
 lines from the \za=0.70152 system. The O~IV $\lambda$ 787.71 line has
  been scaled by a factor of 1.5 for illustration. \label{Fig:AOD7015_1} }
\end{figure}

\begin{figure}
\includegraphics[width=3.5in,height=2.5in]{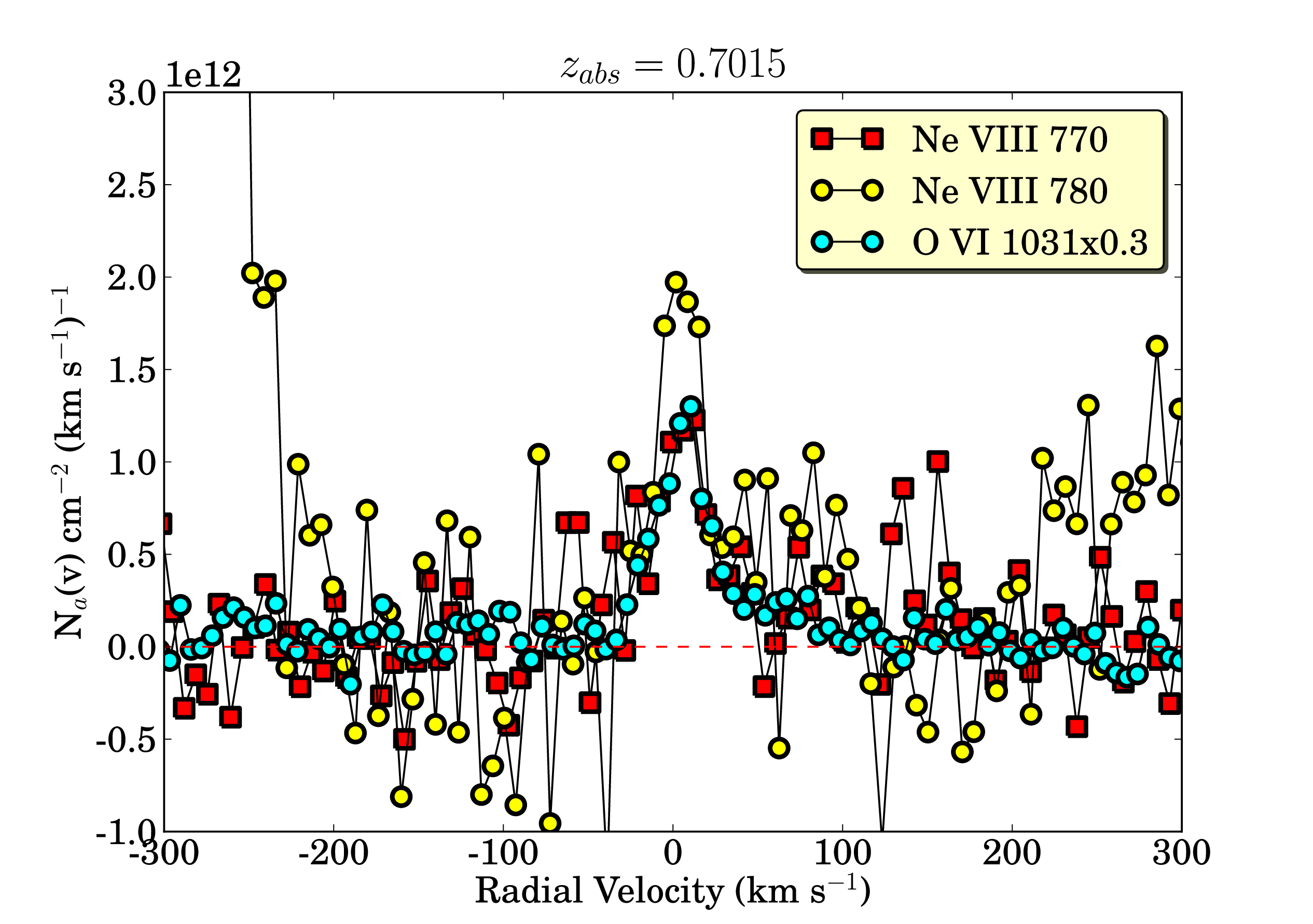} 
\caption{Apparent column density profiles of the O~VI and Ne~VIII
 lines from the \za=0.70152 system. The O~VI $\lambda$ 1031.93 line has
 been scaled by a factor of 0.3 for illustration. \label{Fig:AOD7015_2} }
\end{figure}

\subsection{The \za=0.68381 system:}
The Ne~VIII absorber with the highest affiliated H~I column density
(and the largest number of affiliated metals) is the system at
\za=0.68381. In this absorber we detect lines of H~I, C~III, O~III,
N~IV, O~IV, O~VI, and Ne~VIII.  We also find a marginal feature at the
expected wavelength of the strongest O~II line at 834.47 \AA ;
however, the significance of this feature is $< 3\sigma$, so we can
only place an upper limit on $N$(O~II). Likewise, no statistically
significant N~III lines are evident. Velocity plots for H~I and
metal-line absorption profiles for this system are shown in Figure
\ref{Fig:vel6838}; in this figure and in all velocity plots in this
paper, the velocity scale is in the rest frame of the absorber, and we
define the systemic redshifts of the absorbers to be the centroid of
the strongest component of the O~VI absorption profiles.  Note that in
this figure (and the analogous figures for the other absorbers), we
show some lines to alert the reader to problematic blends with lines
from other redshifts, which are marked with magenta dots. In
Figure~\ref{Fig:vel6838}, the best-fit Voigt profiles are overlaid in
red.  We have used two components to fit the absorption lines in this
system. At this redshift, the H~I \Lya line is redshifted beyond the
long-wavelength end of the COS spectrum. However, we detect H~I
transitions from \Lyb up to \Lye ($\lambda$ 937.80) in the COS FUV
spectrum (see Fig.~\ref{Fig:vel6838}).  Rest-frame equivalent widths and
integrated apparent column densities of the H~I and metal lines in
this system are given in Table \ref{Tab:AOD6838}, and column densities
determined by Voigt profile fitting are listed in Table
\ref{Tab:Cols6838}.

To support our line identifications, we compare the apparent column
density profiles of the O~IV $\lambda$787.71, O~VI \lala 1031.93,
1037.62, and Ne~VIII \lala 770.41, 780.32 lines in Figures
\ref{Fig:AOD6838_1} and \ref{Fig:AOD6838_2}.  While there are multiple
resonance transitions of O~IV in the FUV \citep{Verner94}, at \za =
0.68381, only the O~IV $\lambda$787.71 line is redshifted into the
wavelength range of our COS spectrum.  Therefore, one might question
the reliablility of the O~IV identification.  However, as we can see
from Figures~\ref{Fig:vel6838} and \ref{Fig:AOD6838_1}, the profile of
the of O~IV $\lambda$787.71 line matches the profiles of the O~VI
doublet over a large span of pixels, and this indicates that the O~IV
identification is secure.

Similarly, the $N_{\rm a}(v)$ plots are helpful for the Ne~VIII
identification.  The Ne~VIII $\lambda$780.32 line is certainly blended
with O~III $\lambda$832.93 at $z_{\rm abs}$ = 0.57757, and this
Ne~VIII transition could also be blended with C~III $\lambda$977.02 at
$z_{\rm abs} \approx$ 0.3448.\footnote{There are hundreds of
  absorption lines in the COS spectrum of PG1148+549, and a full
  summary of the lines in the spectrum is beyond the scope of this
  paper.  We plan to analyze other systems in future papers.} The
O~III blend is securely identified; the $z_{\rm abs}$ = 0.57757
absorber shows many H~I Lyman series lines and metals with a
distinctive two-component structure that matches the O~III profile.
One of the O~III components at $z_{\rm abs}$ = 0.57757 is redward of
the expected wavelength of Ne~VIII 780.32 line, but the other O~III
component falls near the center of the expected Ne~VIII feature.  The
contribution of C~III $\lambda$977.02 at $z_{\rm abs} \approx$ 0.3448
to this blend is less certain; a well-detected Ly$\alpha$ line is
found at $z_{\rm abs}$ = 0.3448, and while this Ly$\alpha$ absorber 
has very few affiliated metals, C~III $\lambda$977.02 is one of the
strongest FUV lines, and there could be some optical depth from C~III
in the blend as well.  Given these blends, it is not surprising that
the Ne~VIII $\lambda$ 780.32 profile appears to be stronger than the
Ne~VIII 770.41 line (see Figure~\ref{Fig:AOD6838_2}) because some of
the optical depth in the Ne~VIII 780.32 profile is due to other
species at other redshifts.  Unfortunately, only one O~III line from
$z_{\rm abs}$ = 0.57757 is redshifted into the COS spectrum, so we
cannot remove the O~III optical depth from the blend by modeling.
Therefore, we must rely on the (unblended) Ne~VIII $\lambda$ 770.41
transition alone in the $z_{\rm abs}$ = 0.68381 absorber.  We can find
no other plausible metal-line identifications for the Ne~VIII
$\lambda$ 770.41 candidate, and the fact that the shape of the Ne~VIII
$\lambda$ 770.41 profile matches the shape of the well-detected O~VI
lines at this redshift (see Figures~\ref{Fig:vel6838} and
\ref{Fig:AOD6838_2}) supports the identification of the feature as
Ne~VIII.

Finally, we note that the O III $\lambda$ 832.93 line at $z_{\rm abs}$
= 0.68381 is blended with the Milky Way Si IV $\lambda$ 1402.77
transition. In this case, we can correct for the blend by modeling; by
fitting the corresponding Milky Way Si IV $\lambda$ 1393 line, we can
predict the strength of the Si IV $\lambda$ 1402.77 contamination
(shown with an orange line in Figure~\ref{Fig:vel6838}) and account
for it in our fitting of the O III $\lambda$ 832.93 profile.  A strong
\Lya line at \za=0.1529 is also observed near the O~III $\lambda$
832.93 line (at $-175$ \kms ; see Figure~\ref{Fig:vel6838}). This
Ly$\alpha$ line could contribute some optical depth to the O~III
feature as well due to the extended wings of the COS LSF, so this
Ly$\alpha$ feature was also included in the multiline fit to the O III
at $z_{\rm abs}$ = 0.68381.

\subsection{The \za=0.70152 system}
A variety of absorption lines from metals are evident at \za =
0.70152, including C~III, O~III, O~IV, O~VI, and Ne~VIII. We do not
detect N~III or N~IV, but the only N~IV transition that is redshifted
into the COS spectrum is lost in a blend with the strong Galactic O~I
$\lambda$1302.17 absorption line. The N III $\lambda$ 989.80 line is
blended with a \Lya line at \za=0.3862, but the N III 685.51
transition, which is nearly two times stronger, is not observed
either. As in the \za = 0.68381 system, the H~I Ly$\alpha$ line is
redshifted out of the COS spectrum, but unlike the \za =0.68381
absorber, no higher Lyman series lines are detected in the \za=0.70152
system.  Velocity plots of the absorption lines from this system,
including the undetected Ly$\beta$ line, are shown in Figure
\ref{Fig:vel7015}, with the best fit profiles overlayed in red.
Equivalent widths and column densities for the \za = 0.70152 absorber,
measured with the apparent column density technique and profile
fitting, can be found in Tables~\ref{Tab:AOD7015} and
\ref{Tab:Cols7015}.  Two well-detected components at $v = 0$ and 59 km
s$^{-1}$ are readily apparent in the absorption profiles of C~III,
O~IV, and O~VI, but Ne~VIII is clearly detected only in the $v = 0$ km
s$^{-1}$ feature.  A third component is clearly seen at $v = -125$ km
s$^{-1}$ in the C~III and O~IV transitions, and weak but consistent
features are present at this velocity in O~III and O~VI.  The O~VI at
this velocity is particularly marginal; the feature is only seen in
the $\lambda 1031.93$ line and is not confirmed by the (albeit noisy)
$\lambda 1037.62$ profile.  In this paper, we focus on the two
components at $v = 0$ and 59 km s$^{-1}$ that are well-detected in
O~VI.  Higher S/N would be helpful for the study of weaker features
such as the $v = -125$ km s$^{-1}$ component.

\begin{table}
\caption{Equivalent Widths and Integrated Apparent Column Densities for Transitions Observed in the \za=0.70152 System$^{a}$ \label{Tab:AOD7015}}
\begin{center} 
\begin{tabular}{lcc}
\hline\hline
  Transition               &            $W_{\rm r}$ (m\AA )     &   log [$N_{\rm a}$(cm $^{-2}$)]      \\
\hline
H I $\lambda$1025.72\dotfill        &         $< 36^{\rm b}$     &   $<13.7^{\rm c}$     \\
C~II $\lambda$903.96\dotfill        & $<18^{\rm b}$              & $<12.9^{\rm c}$ \\
N~II $\lambda$915.61\dotfill        & $<31^{\rm b}$              & $<13.4^{\rm c}$ \\
O~II $\lambda$834.47\dotfill        & $< 20^{\rm b}$             & $<13.4^{\rm c}$ \\
Mg~II $\lambda$2796.35\dotfill      & $< 17^{\rm b,d}$           & $<11.6^{\rm c}$ \\
C III $\lambda$977.02\dotfill       &       126$\pm$16           &        13.41$\pm$0.05             \\
N III $\lambda$685.51\dotfill       & $< 26^{\rm b}$             & $<13.4^{\rm c}$ \\
O III $\lambda$832.93\dotfill       &        18$\pm$5            &     13.48$\pm$0.09             \\
S~III $\lambda$698.73\dotfill       & $<13^{\rm b}$              & $<12.6^{\rm c}$   \\
O IV $\lambda$787.71\dotfill        &     75$\pm$9               &       14.18$\pm$0.04            \\
S~IV $\lambda$809.67\dotfill        & $<19^{\rm b}$              & $<13.5^{\rm c}$   \\
S~V $\lambda$786.48\dotfill         & $<20^{\rm b}$              & $<12.4^{\rm c}$ \\
O VI $\lambda$1031.93\dotfill       &     168$\pm$19             &       14.29$\pm$0.04            \\
O VI $\lambda$1037.62\dotfill       &      113$\pm$19            &       14.35$\pm$0.06             \\
S~VI $\lambda$933.38\dotfill         & $<30^{\rm b}$              & $<12.9^{\rm c}$ \\
Ne VIII $\lambda$770.41\dotfill     &     28$\pm$5               &       13.75$\pm$0.07               \\
Ne VIII $\lambda$780.32\dotfill     &     18$\pm$6               &       13.86$\pm$0.11               \\
\hline
\end{tabular}
\end{center}
$^{\rm a}$ Quantities in this table do not include the weak component at $v = -125$ km s$^{-1}$, which is only marginally detected in O~VI and is not detected in Ne~VIII. \\
$^{\rm b}$ Three $\sigma$ upper limit obtained by integration over $-50 < v < 100$ km s$^{-1}$. \\
$^{\rm c}$ Upper limit based on the $3\sigma$ upper limit on $W_{\rm r}$, assuming the line is in the linear regime of the curve of growth. \\
$^{\rm d}$ Derived from the Keck/HIRES data.  
\end{table}

\begin{table}
\caption{Column Densities from Profile Fits to the COS Data for the \za=0.70152 System \label{Tab:Cols7015}}
\begin{center}
\begin{tabular}{lrcccccccccccc}
\hline\hline
   Ion           & $v_{rad}$ \ \ \ \  & log [N (cm$^{-2}$)] &  $b$      \\
                 & (\kms\ )           &                     &  (\kms\ )           \\
\hline
C III\dotfill    & $-125$   \ \ \ \   & 13.24$\pm$0.24      & 6.3$\pm$3.2               \\
                 & 0     \ \ \ \     & 13.39$\pm$0.03      & 23.9$\pm$2.2               \\
                 & 59   \ \ \ \      & 12.63$\pm$0.10      & 18.7$\pm$6.9               \\
O III\dotfill    & $-125$    \ \ \ \  & 13.39$\pm$0.07      & 20.4$\pm$7.7                 \\
                 & 0        \ \ \ \  & 13.55$\pm$0.06      & 23.9$\pm$5.5                 \\
                 & 59     \ \ \ \    &  --                 &    --                               \\
O IV\dotfill     & $-125$ \ \ \ \     & 13.63$\pm$0.06      & 14.8$\pm$4.6                 \\
                 & 0   \ \ \ \       & 14.17$\pm$0.02      & 23.7$\pm$1.9                 \\
                 & 59$^{a}$ \ \ \    & 13.33$\pm$0.15$^{\rm a}$& 42.7$\pm$21.0$^{\rm a}$        \\
O VI\dotfill     & $-125^{b}$ \ \ \   & 13.46$\pm$0.14$^{\rm b}$& 32.4$\pm$19.1$^{\rm b}$           \\
                 & 0     \ \ \ \     & 14.28$\pm$0.03      & 20.9$\pm$2.0                  \\
                 & 59    \ \ \ \     & 13.63$\pm$0.12      & 27.3$\pm$10.3              \\
Ne VIII\dotfill  & $-125$  \ \ \ \    &  --                 &  --                         \\
                 & 0      \ \ \ \    & 13.82$\pm$0.06$^{\rm c}$& 28.2$\pm$7.1$^{\rm c}$          \\
                 & 59   \ \ \ \      &  --                 &     --                                    \\                                                                                                                                                                       
\hline
\end{tabular}
\end{center}
$^{\rm a}$ While O~IV absorption is clearly present at this velocity, the parameters of this component are more uncertain due to the intrinsic weakness of the component and its significant blending with the stronger adjacent component at $v = 0$ km s$^{-1}$. \\
$^{\rm b}$ This component is marginally detected in the stronger $\lambda$1031.93 transition and is not adequately detected in the weaker $\lambda$1037.62 line; consequently, the parameters of the component are highly uncertain. \\
$^{\rm c}$ The Ne~VIII $\lambda$780.32 line is affected by a blend (see text), so the results in this table are based on a fit to the Ne~VIII $\lambda$770.41 transition only. \\
\end{table}

As in the previous section, we also compare the $N_{\rm a}$ profiles
of O~IV, O~VI, and Ne~VIII in Figures~\ref{Fig:AOD7015_1} and
\ref{Fig:AOD7015_2} to aid our comments about line identification. We
can see from Figure~\ref{Fig:AOD7015_1} that the O~IV $\lambda$787.71
identification is unambiguous; the O~IV and O~VI profiles match
precisely, including an asymmetric wing extending to positive
velocities.  We can see that this wing is due to a weak component at
$v = $ +60 km s$^{-1}$, and we fit the profiles with two components at
$v = $ 0 and +60 km s$^{-1}$.  The Ne~VIII $\lambda$780.32 profile
shows some excess optical depth compared to the Ne~VIII
$\lambda$770.41 line (Figure~\ref{Fig:AOD7015_2}).  This could be due
to noise, but we note that there appears to be some additional excess
absorption on the red side of the 780.32 \AA\ line, and it possible
that the Ne~VIII $\lambda$780.32 profile is weakly contaminated by an
interloper, likely a weak Ly$\alpha$ line from lower
redshift. However, the good match of the Ne~VIII $\lambda$770.41
profile to the O~IV and O~VI profiles indicates that reliable Ne~VIII
measurements can be obtained from the 770.41 \AA\ data.

The absence of Ly$\beta$ absorption is an interesting feature of this
absorber -- combined with detection of strong metal lines, this
suggests that this is a highly metal enriched system, as we will
discuss further in \S \ref{Sec:Ion}.  Integrating over the full
velocity range of the detected metal absorption ($-50 < v < 100$ km
s$^{-1}$), we obtain log $N$(H~I) $< 13.7 \ (3\sigma)$. However, we
noted above that two components are evident in this system, and will
consider the implied metallicities of this absorber on a
component-by-component basis in the following section, so have also
obtained limits on $N$(H~I) in each of the components.  For the
feature at $v = 0$ km s$^{-1}$, we integrate from $-50$ to 50 km s$^{-1}$
and obtain log $N$(H~I) $< 13.6 \ (3\sigma)$.  For the $v = 60$ km
s$^{-1}$ component, we integrate from 50 to 100 km s$^{-1}$
and obtain log $N$(H~I) $< 13.4 \ (3\sigma)$. 

\subsection{The \za=0.72478 system}

\begin{figure*}
\plotone{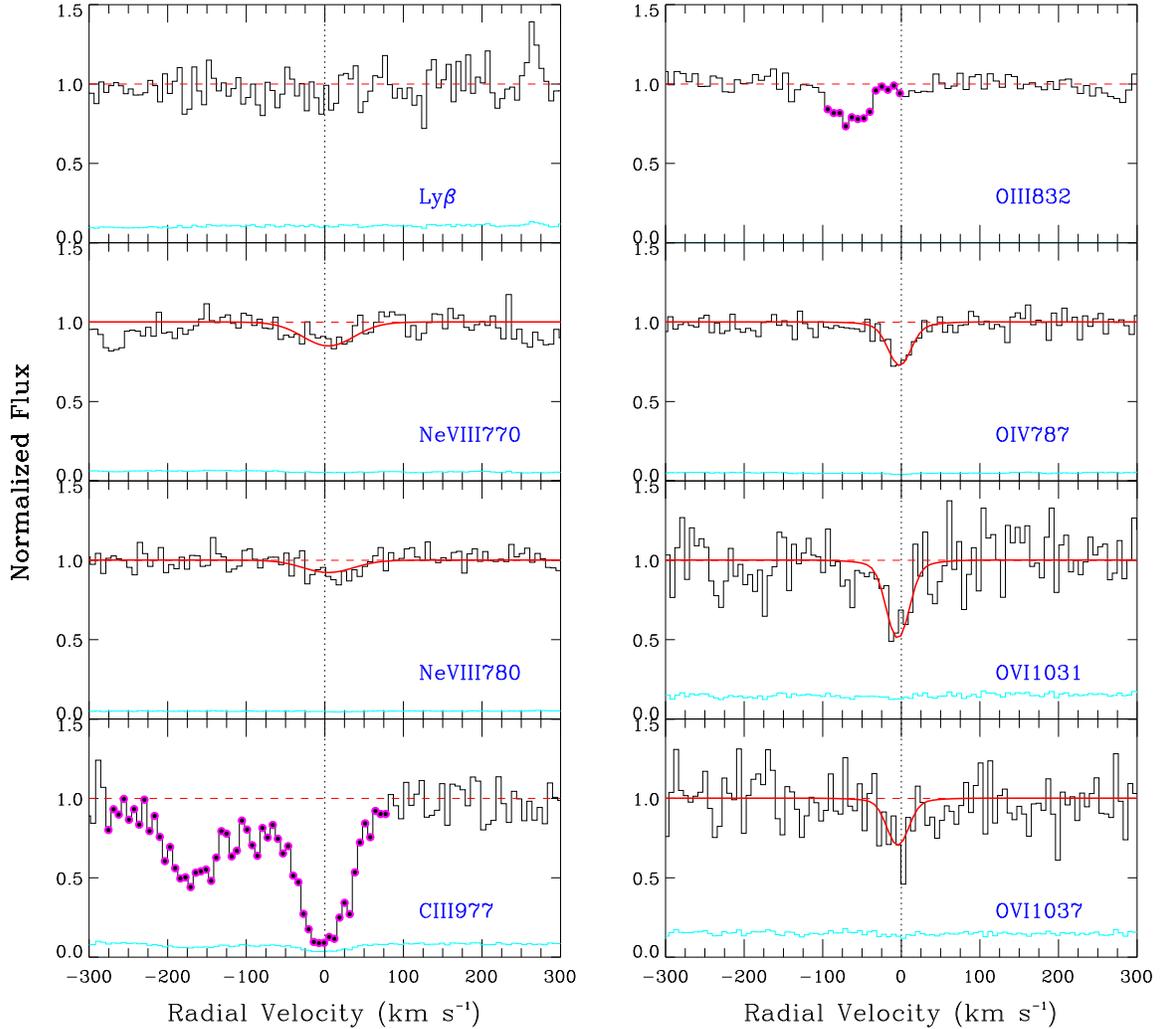}
\caption{Continuum-normalized absorption profiles of the \za=0.72478
  absorption system plotted in the rest frame of the absorber, as in
  Figure~\ref{Fig:vel6838}.  \label{Fig:vel7248} }
\end{figure*}

\begin{figure}
\includegraphics[width=3.5in,height=2.5in]{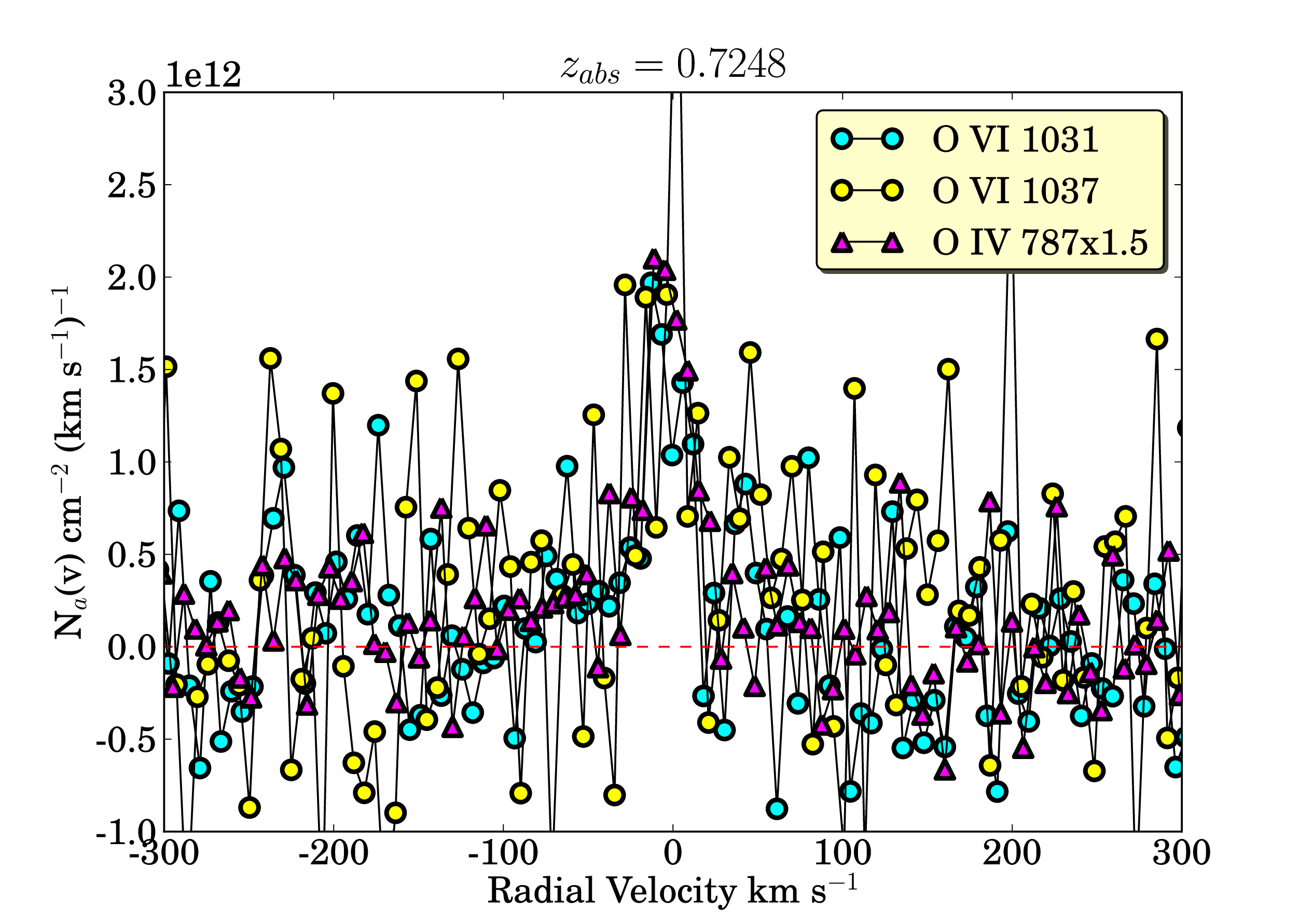}
\caption{Apparent column density profiles of the O~IV and O~VI lines
  of the \za=0.72478 system. The O~IV $\lambda$ 787.71 line has been
  scaled by a factor of 1.5 for purposes of
  comparison. \label{Fig:AOD7248_1} }
\end{figure}

\begin{figure}
\includegraphics[width=3.5in,height=2.5in]{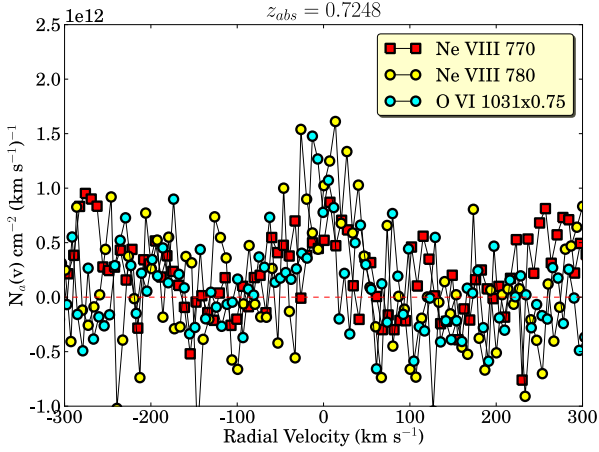}  
\caption{Apparent column density profiles of the O~VI and Ne~VIII
  lines from the \za=0.72478 system.  The O~VI $\lambda$ 1031.93 line
  has been scaled by a factor of 0.75 for comparison with the other
  lines. \label{Fig:AOD7248_2} }
\end{figure}

The final Ne~VIII absorber is similar to the system at \za=0.70152
with detections of O~IV, O~VI, and Ne~VIII and no affiliated H~I
absorption in Ly$\beta$ or higher Lyman series lines.  
We show a velocity stack plot of selected lines in
Figure~\ref{Fig:vel7248}, and we compare the $N_{\rm a} (v)$ profiles
of O~IV, O~VI, and Ne~VIII in Figures \ref{Fig:AOD7248_1} and
\ref{Fig:AOD7248_2}. The data indicate only a single component in this
absorber. No H~I absorption is detected, and we derive an upper limit
of log $N$(H~I) $<$13.7 by integrating the \Lyb $\lambda$ 1025 line
region from $-75$ to +75 km s$^{-1}$. Equivalent widths and column
densities for this system are summarized in Tables \ref{Tab:AOD7248}
and \ref{Tab:Cols7248}. The O~IV $\lambda$ 787.71 and O~VI \lala
1031.93, 1037.62 lines are detected in this system at high
significance, and the good correspondence of the Ne~VIII profiles with
the other lines supports the Ne~VIII identification.

\begin{table}
\caption{Equivalent Widths and Integrated Apparent Column Densities for Transitions Observed in the \za=0.72478 System. 
                \label{Tab:AOD7248}}
\begin{center}
\begin{tabular}{lcc}
\hline\hline
  Transition                    & $W_{\rm r}$ (m\AA )   &   log [$N_{\rm a}$ (cm$^{-2}$)] \\
\hline
H I $\lambda$1025.72\dotfill    &      $<30^{\rm a}$    & $<13.7^{\rm b}$   \\
C~II $\lambda$687.05\dotfill    & $<13^{\rm a}$         & $<12.9^{\rm b}$ \\
N~II $\lambda$671.41\dotfill    & $<23^{\rm a}$         & $<13.9^{\rm b}$ \\
O II $\lambda$834.47\dotfill    &       $<15^{\rm a}$   & $<13.5^{\rm b}$   \\
Mg~II $\lambda$2796.35\dotfill  & $<16^{\rm a,c}$       & $<11.6^{\rm b}$ \\
N~III $\lambda$685.00$^{\rm d}$\dotfill   & $<14^{\rm a}$ & $<13.4^{\rm b}$ \\
O III $\lambda$832.93\dotfill   &     $<16^{\rm a}$     & $<13.5^{\rm b}$                            \\
S~III $\lambda$677.75\dotfill   & $<21^{\rm a}$         & $<12.5^{\rm b}$ \\
N IV $\lambda$765.15\dotfill    & $<26^{\rm a}$         & $<12.9^{\rm b}$                                \\
O IV $\lambda$787.71\dotfill    &        27$\pm$5       &    13.70$\pm$0.06                \\
S~IV $\lambda$748.40\dotfill    & $<13^{\rm a}$         & $<12.7^{\rm b}$ \\
S~V $\lambda$786.48\dotfill     & $<14^{\rm a}$         & $<12.2^{\rm b}$ \\
O VI $\lambda$1031.93\dotfill   &       71$\pm$20       &  13.84$\pm$0.10                \\
O VI $\lambda$1037.62\dotfill   &       37$\pm$12       &   13.87$\pm$0.15                \\
S~VI $\lambda$933.38\dotfill    & $<30^{\rm a}$         & $<12.9^{\rm b}$ \\
Ne VIII $\lambda$770.41\dotfill &        26$\pm$8       &       13.70$\pm$0.12                                   \\
Ne VIII $\lambda$780.32\dotfill &       19$\pm$5        &       13.87$\pm$0.10                               \\
\hline
\end{tabular}
\end{center}
$^{\rm a}$ Three $\sigma$ upper limit obtained by integration over $-75 < v < 75$ km s$^{-1}$. \\
$^{\rm b}$ Upper limit based on the $3\sigma$ upper limit on $W_{\rm r}$, assuming the line is in the linear regime of the curve of growth. \\
$^{\rm c}$ Derived from the Keck/HIRES data. \\
$^{\rm d}$ We use the N~III $\lambda$685.0 transition to place this limit because the other N~III lines in the COS spectrum at this redshift are confused by blending. \\
\end{table}

\begin{table}
\caption{Column Densities from Profile Fits to the COS Data for the \za=0.72478 System. \label{Tab:Cols7248}}
\centering 
\begin{tabular}{lrccccccccccccc}
\hline\hline
Ion \ \ \ \ \ \ \  & $v_{rad}$ \ \ \ & log [N (cm$^{-2}$)] & $b$      \\
                   & (km s$^{-1}$)   &                     & (km s$^{-1}$) \\
\hline
O IV\dotfill       & 0  \ \ \ \ \    & 13.75$\pm$0.04      & 16.1$\pm$2.9  \\ 
O VI\dotfill       & 0  \ \ \ \ \    & 13.86$\pm$0.07      & 15.8$\pm$3.8  \\
NeVIII\dotfill     & 0  \ \ \ \ \    & 13.81$\pm$0.06      & 41.4$\pm$7.5  \\
\hline
\end{tabular}
\end{table}

\section{Ionization Modeling} 
\label{Sec:Ion}

We now turn to analysis of the physical conditions and metallicity of
the Ne~VIII/O~VI absorbers.  Given the velocity alignment of the
Ne~VIII and O~VI absorption lines with lower ionization stages, it is
clear that there is some sort of relationship between the
Ne~VIII-bearing gas and the lower ionization material.  The velocity
alignment does not necessarily indicate that the Ne~VIII arises in a
single-phase cloud that is cospatial with the lower ions; it is
possible that this indicates that the Ne~VIII originates in an
interface on the surface of a lower ionization cloud, for example. To
investigate this relationship and probe the nature of the Ne~VIII
absorbers, we holistically analyze the ionization mechanism(s) for
{\it all} detected ionization stages, not just the Ne~VIII.  For this
purpose, we follow the methodology of \citet{Tripp11}, which is
briefly summarized as follows.  We first constrain the portion of the
absorbing gas that can be attributed to low-density, cool gas that is
photoionized by the UV background light (\S~\ref{Sec:Photoion}).
Photoionization models of optically thin gas homologously depend on
the ionization parameter $U$, which is the ratio of the number density
of ionizing photons to the particle density ($U=n_{\gamma}/n_{H}$),
the absolute and relative abundances of the elements, and the hydrogen
column density.  In order to avoid modeling ambiguities resulting from
degeneracies between ionization corrections and relative abundances,
we use the O~IV/O~III column-density ratio to determine $U$, and we
constrain the models to match the measurements of (or upper limits on)
$N$(H~I).  With the ionization parameter constrained by O~IV/O~III, we
then consider the column densities predicted by the photoionization
models for the other species.  We will find that
the photoionization models predict O~VI and Ne~VIII column densities
that are orders of magnitude lower than the observed $N$(O~VI) and
$N$(Ne~VIII) values. Consequently, we then examine whether the O~VI
and Ne~VIII can be attributed to a collisionally ionized hot phase
(\S~\ref{Sec:Collion}).

\subsection{Photoionization}
\label{Sec:Photoion}

\begin{figure}
\includegraphics[width=3.3in, angle=0]{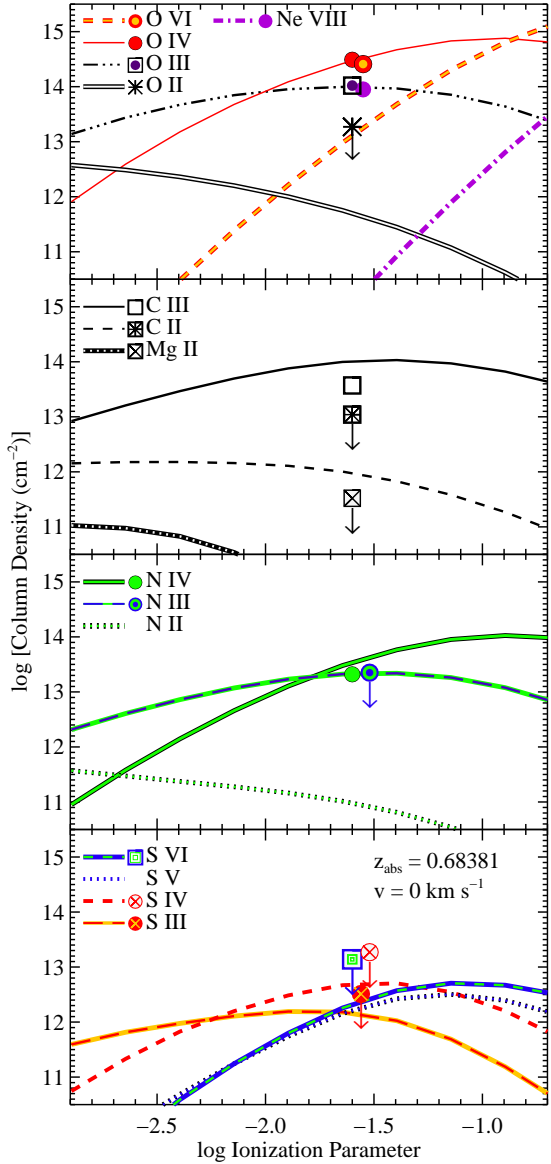}
\caption{Photoionization modeling of the $v = 0$ km s$^{-1}$ component
  in the multiphase O~VI/Ne~VIII absorber at $z_{\rm abs}$ = 0.68381,
  assuming the gas is photoionized by the diffuse UV background from
  quasars and galaxies as calculated by \citet{HM01}.  In each panel,
  the column densities predicted by the photoionization model, as a
  function of ionization parameter $U$, are shown with smooth
  curves. The observed column densities and $3\sigma$ upper limits are
  indicated with large symbols at log $U \approx -1.6$.  Generally,
  the uncertainties in the observed column densities are comparable to
  the symbol sizes (see Tables~\ref{Tab:AOD6838} --
  \ref{Tab:Cols7248}).  The species corresponding to each curve and
  symbol are indicated by the legend in the upper-left corner of each
  panel. In this figure, the overall metallicity is $Z = 0.3
  Z_{\odot}$. \label{Fig:Ionization} }
\end{figure}

We model photoionized gas using \textsc{cloudy} version 8.00
\citep{Fer98}. We have used the Haardt $\&$ Madau extragalactic
background spectrum \citep{HM96} at $z = 0.7$, with the modifications
and updates described in \citet{HM01}, including contributions from
both active galactic nuclei (AGN) and galaxies.  We treat the
absorbers as plane-parallel slabs with constant density, and we
initially assume that the elements have the same relative abundances
observed in the Sun \citep{Lodd03,Asp09}; after running the initial
models, we consider whether there is evidence for departures from the
solar abundance patters.  Local flux sources (starlight from galaxies)
can be brighter than the diffuse UV background if the escape fraction
of the ionizing flux is sufficiently high \citep{Fox05, Misawa09}.
However, observational studies have not yet provided strong evidence
of a high escape fraction, so we will mainly employ the \citet{HM96}
flux as our fiducial model in this paper.

\subsubsection{$z_{\rm abs}$ = 0.68381}

Figure~\ref{Fig:Ionization} shows our photoionization model for the
component at $v$ = 0 km s$^{-1}$ of the absorption system at $z_{\rm
  abs}$ = 0.68381.  In this figure, the column densities predicted by
the \textsc{cloudy} model are plotted with smooth curves, and the
observed column densities and upper limits are displayed with large,
discrete symbols at the value of $U$ that matches the observed
O~IV/O~III ratio (in some cases, small offsets have been applied for
clarity because the symbols overlap).  The legends in each panel label
the curves and symbols.  This model matches the observed O~IV/O~III
ratio with log $U$ = $-1.60$ and metallicity\footnote{We use the usual
  notation where $Z$ is the absolute metallicity and the logarithmic
  metallicity [X/H] = log [$N$(X)/$N$(H)] - log (X/H)$_{\odot}$.}
[X/H] = $-0.5$. Various parameters of this photoionization model are
summarized in Table~\ref{Tab:Photo}.

With the ionization parameter constrained by O~IV/O~III, several
interesting implications unfold.  First and foremost, this photoionization
model falls woefully short of the observed column densities of O~VI
and Ne~VIII: the observed $N$(O~VI) is 25 times higher than the model
prediction, and the observed $N$(Ne~VIII) is many orders of magnitude
higher than the model, which predicts negligible amounts of Ne~VIII in
the photoionized gas at log $U = -1.6$. The high-ion absorption must
arise in a separate phase.  Second, the model is roughly consistent
with the measurements for intermediate-ionization stages of carbon,
nitrogen, and sulfur (lower panels).  However, the model seems to
require modest underabundances of carbon and nitrogen; the observed
columns of both C~III and N~IV are somewhat lower than predicted by
the model, and the upper limit on N~III is also in tension with the
model, which indicates that N~III should be marginally
detected.\footnote{We note that a weak feature is apparent at the
  expected wavelength of N~III $\lambda$685.51, but the feature is not
  recorded at adequate significance to provide a reliable
  measurement.} To fit the C~III and N~IV, this model requires modest
relative abundance adjustments; relative to oxygen, the model requires
[C/O] = $-0.4$ and [N/O] = $-0.2$.  The upper limits on the sulfur
ions are consistent with the model but do not provide additional
information.  The sulfur ion column densities are all expected to be
below the detection threshold (lowest panel in
Figure~\ref{Fig:Ionization}).  However, the predicted sulfur columns
are only moderately below the observed upper limits, so higher
signal-to-noise observations could provide information on the relative
abundances of sulfur as well.  Even though the Mg~II $\lambda \lambda$
2796.35, 2803.53 doublet is extremely strong and places tight upper
limits on $N$(Mg~II), we see from Figure~\ref{Fig:Ionization} that
Mg~II is easily photoionized in these conditions and is expected to be
undetectable.  Indeed, the photoionization model indicates that all of
the low ions that we can constrain (e.g., C~II, N~II, O~II, Mg~II)
should have very low column densities that are well below the observed
upper limits.  As we discuss below, underabundances of C and N are not
necessarily surprising.

We note that the discrepancies between the model and the observed O~VI
and Ne~VIII columns, as well as the underabundances implied by C~III
and N~IV, cannot be attributed to measurement uncertainties; the
column density uncertainties are generally smaller than the symbol
sizes in Figure~\ref{Fig:Ionization}.  A potential source of
systematic uncertainty in this analysis is the assumed ionizing flux
field. The shape and intensity of the UV background are constrained by
models \citep[e.g.,][]{HM96} but are only loosely constrained by
observations \citep[][and references therein]{DT01}.  In principle,
with fine tuning, modifications of the shape of the radiation field
could alleviate the C~III and N~IV discrepancies, but usually this is
not helpful because the ionization potentials of O~III/O~IV,
C~III/C~IV, and N~III/N~IV are too similar.  The observations may be
indicating real underabundances of C and N.

\begin{table*}
\footnotesize
\begin{center}
\caption{Photoionization Model Parameters$^{\rm a}$\label{Tab:Photo}}
\begin{tabular}{llcccrccll}
\hline\hline
Redshift \ \ \  & Component    & log N(H I)& log N(H$_{tot}$)& log $U^{\rm b}$ & [X/H]$^{\rm c}$  & [C/O]$^{\rm d}$ & [N/O]$^{\rm d}$ & Number      & Absorber \\
                & Velocity     &       \  &      \           &                 &                  &   \             & \               & Density     & Thickness \\
   \            & (km s$^{-1}$)&       \  &      \           &   \             &  \               &   \             & \               & (cm$^{-3}$) & \ \ \ (kpc) \\
\hline
0.68381\dotfill &           0             & 14.65    &  18.49    & $-1.60$ & $-0.5$   & $-0.4$ & $-0.2$ & 2.9$\times10^{-4}$   & \ \ \ \ 3.5 \\
0.68381\dotfill &           40            & 14.24    &  18.24    & $-1.45$ & $-0.5$   & $-0.5$ & $-0.4$ & 2.0$\times10^{-4}$   & \ \ \ \ 2.8 \\
0.70152\dotfill &           0             & $<$13.60 &  $<$17.54 & $-1.38$ & $>0.2$   & $-0.1$ & --     & 1.8$\times10^{-4}$   & \ \ $<$0.6 \\
0.70152\dotfill & 59 (Model 1)$^{\rm e}$  & $<$13.40 &  $<$17.11 & $-1.67$ & $> -0.2$ & $-0.3$ & --     & 3.5$\times10^{-4}$   & \ \ $<$0.1 \\
0.70152\dotfill & 59 (Model 2)$^{\rm e}$  & $<$13.40 &  $<$18.31 & $-0.70$ & $> -0.8$ & $+0.5$ & --     & 3.7$\times10^{-5}$   & \ \ $<$18  \\
0.72478\dotfill & 0 \ (Model 1)$^{\rm e}$ & $<$13.70 &  $<$17.26 & $-1.75$ & $>0.0$   & --     & --     & 4.4$\times10^{-4}$   & \ \ $<$0.1 \\
0.72478\dotfill & 0 \ (Model 2)$^{\rm e}$ & $<$13.70 &  $<$18.60 & $-0.70$ & $> -0.7$ & --     & --     & 3.9$\times10^{-5}$   & \ \ $<$33 \\
\hline
\end{tabular}
\end{center}
$^{\rm a}$ Photoionization modeling results, assuming the gas is
photoionized by the UV background flux from QSOs and galaxies, as
calculated by \citet{HM96} with the updates of \citet{HM01}.
Plausible modification of the ionizing flux field can systematically
change photoionization model results by factors of $\sim 2$
\citep{Aracil06,Tripp06,Stocke13}. \\
$^{\rm b}$ $U$ = ionization parameter. \\
$^{\rm c}$ Logarithmic absolute abundance with respect to solar abundances, based on oxygen. \\
$^{\rm d}$ Logarithmic relative abundance of carbon and nitrogen with respect to their solar abundances compared to oxygen. \\
$^{\rm e}$ For components in which O~III absorption is not detected, we consider two models.  In model 1, the lower limit on the O~IV/O~III column-density ratio is used to find the minimum ionization parameter that is consistent with the O~III and O~IV constraints.  In model 2, the measured column densities of O~IV and O~VI (which are both detected) are used to find the maximum ionization parameter allowed by the measurements. As discussed in the text, model 1 is strongly favored over model 2 in these cases, and at any rate, none of the photoionization models produce enough Ne~VIII to match the observations in any of these three absorption systems. \\
\end{table*}

On the other hand, it is interesting to consider whether adjustments
of the ionizing flux field could lead to higher O~VI and Ne~VIII
columns and thereby alleviate this problem.  For example, it is known
that obscured AGN comprise an important portion of the X-ray
background \citep[e.g.,][]{Mushotzky2000,Gilli07}; could an obscured
AGN near the PG1148+549 sight line change the ionizing flux field in a
way that boosts the O~VI and Ne~VIII column densities?  To explore
this idea, we have used \textsc{cloudy} to produce models of an
obscured AGN with obscuring columns as high as log $N$(H) = 23.  Our
absorbers are optically thin are thus are always exposed to the UV
background, so we combined the flux from the obscured AGN with the
Haardt \& Madau UV background with various mixtures of relative
brightness of the two components of the model.  Every variation of
this obscured AGN + UV background model failed badly.  As can be seen
in, e.g., Figure 1d of \citet{Hamann97}, obscuration of an AGN actually
reduces the flux of photons capable of photoionizing Ne~VII to produce
Ne~VIII more than it reduces the flux that can ionize O~II, O~III,
O~IV, etc., so adding an obscured AGN does not boost the Ne~VIII or
O~VI column densities relative to the lower ionization stages.

It appears that the Ne~VIII and O~VI must arise in a separate gas
phase from the lower ionization stages.  But is the Ne~VIII-bearing
gas photoionized or collisionally ionized?  We can rule out
photoionized Ne~VIII and O~VI -- our \textsc{cloudy} models indicate
that in order to arise in photoionized gas, the observed Ne~VIII and
O~VI columns would require very high values for $U$. With the Haardt
\& Madau flux at $z = 0.68$, this in turn requires very low densities
and cloud sizes $>1$ Mpc. Such large sizes would result in broadening
of the lines by the Hubble flow that is not allowed by the measured
line widths. Similar arguments have been made for other systems with
Ne VIII detections, where the implied cloud sizes in the photoionized
Ne~VIII models leads to unrealistically large cloud sizes
\citep{Sav05,Nar11,Tripp11,Nara12}.

As single-phase and two-phase photoionization models fail to account
for both the low and high ions simultaneously, we conclude that the
lower ions (O~III, C~III, O~IV, etc.) reside in a photoionized core,
and that the high ions (O~VI and Ne~VIII) reside in a collisionally
ionized, hot phase. Such necessity of a collisionally ionized phase
has also been demonstrated for other Ne VIII absorbers \citep{Nar09,
  Nar11, Nara12, Sav05, Tripp11}. We will investigate the implied
properties of the hot phase in \S~\ref{Sec:Collion}.

We have applied the same photoionization models to the component at $v
= 40$ km s$^{-1}$ in the $z_{\rm abs}$ = 0.68381 absorption system.
As summarized in Table~\ref{Tab:Photo}, we obtain very similar results
as we found for the $v = 0$ km s$^{-1}$ component, including the same
overall metallicity ($Z = 0.3 Z_{\odot}$), similar underabundances of
carbon and nitrogen (relative to oxygen), and a similar ionization
parameter.

\subsubsection{$z_{\rm abs}$ = 0.70152 and 0.72478}

We have used our \textsc{cloudy} photoionization models to similarly
constrain the properties of the systems at $z_{\rm abs}$ = 0.70152 and
0.72478, and our results are listed in Table~\ref{Tab:Photo}.  For
these absorption systems, we only have upper limits on the H~I column
density and we detect a more limited suite of metals, but we can,
nevertheless, show that the physical conditions and metallicity are
similar to those of the $z_{\rm abs}$ = 0.68381 system.

For the $v = 0$ km s$^{-1}$ component of the $z_{\rm abs}$ = 0.70152
absorber, the observed O~IV/O~III ratio indicates that log $U = -1.38$
and, interestingly, the observed columns require a supersolar
metallicity, [X/H] $> +0.2$.  We can only place a lower limit on the
metallicity since we only have an upper limit on $N$(H~I); a decrease
in the $N$(H~I) assumed for the photoionization model [log $N$(H~I)
  $\leq$ 13.60] would require an increase in the metallicity.
However, in this component the evidence for a carbon underabundance is
less compelling with [C/O] = $-0.1$, and we have no information about
nitrogen.  On the other hand, we once again draw a robust conclusion
about the highly ionized gas: the O~VI and Ne~VIII column densities
are orders of magnitude higher than expected in photoionization models
-- these models cannot simultaneously match O~III, O~IV, O~VI, and
Ne~VIII in a single-phase gas cloud.

Due to blending, we are unable to measure $N$(O~III) in the $v = 59$
km s$^{-1}$ component of the $z_{\rm abs}$ = 0.70152 system (see
Figure~\ref{Fig:vel7015}), which leads to greater ambiguity in the
modeling.  We bracket the physical conditions and metallicity in this
component by considering two photoionization models that are at the
extreme ends of the ionization parameter range allowed by the data.
These models are referred to as Model 1 and Model 2 in
Table~\ref{Tab:Photo}.  In Model 1, we find the ionization parameter
that matches the observed minimum O~IV/O~III ratio [based on the
  measured $N$(O~IV) and the upper limit on $N$(O~III)]. As in the
previous section, this model falls far short of the observed
$N$(O~VI).  In Model 2, we find the value of $U$ that matches the
observed O~IV/O~VI ratio.  By definition, Model 2 agrees with the
measured $N$(O~VI), but we note that this model does not produce
enough Ne~VIII compared to the observations.

While it only provides a lower limit on $U$, we argue that Model 1 is
likely closer to the actual physical conditions and metallicity of the
gas for the following reasons. First, Model 2 does not predict enough
C~III compared to the measured $N$(C~III), and in order to fit the
data, we must invoke a substantial carbon overabundance.  There is
astrophysical precedence for a carbon \textsl{under}abundance --
carbon underabundances are observed in stars in the disk and halo of
the Milky Way \citep{Ak04,Bensby06, Fabbian09} as well as damped Ly$\alpha$
absorbers \citep{Pet08,Penprase10,Cooke11}. On the other hand, the
combination of low metallicity and a high carbon overabundance
required by Model 2, [X/H] = $-0.8$ and [C/O] = +0.5, is a rather
peculiar abundance pattern.  There are some ``carbon-enhanced'' damped
Ly$\alpha$ systems \citep{Cooke12}, but these absorbers have much
lower metallicities than Model 2.  It seems likely that Model 2
requires a carbon overabundance because the model is wrong.  Second,
as shown in Table~\ref{Tab:Photo}, Model 1 implies absorber properties
(e.g., density and size) that are similar to the characteristics of
the other components where O~III \textit{is} detected, whereas Model 2
requires much lower densities and much larger gas clouds.  Third, if
we generally place the O~VI in the same (cospatial) gas as the O~IV,
then it becomes hard to explain where the Ne~VIII arises (Ne~VIII must
then orginate in a phase that makes no contribution to the O~VI column
density). If we place the O~VI in the Ne~VIII-bearing gas, which is
distinct from the O~IV phase, then the Ne~VIII is naturally explained,
as we discuss below.  At any rate, Models 1 and 2 bracket the allowed
range of physical conditions, as summarized in Table~\ref{Tab:Photo}.

The Ne~VIII absorber at $z_{\rm abs}$ = 0.72478 is very similar to the
absorption system at $z_{\rm abs}$ = 0.70152, and we have modeled the
$z_{\rm abs}$ = 0.72478 system in the same way as we treated the
absorber at $z_{\rm abs}$ = 0.70152.  Our results for the $z_{\rm
  abs}$ = 0.72478 case are summarized in Table~\ref{Tab:Photo}.  As in
the other systems studied in this paper, the $z_{\rm abs}$ =0.72478
requires a high metallicity, as one might expect given the detection
of metals and the absence of H~I (Figure~\ref{Fig:vel7248}), and the
presence of Ne~VIII requires a hot, collisionally ionized phase.

\begin{figure*}
\includegraphics[width=17cm]{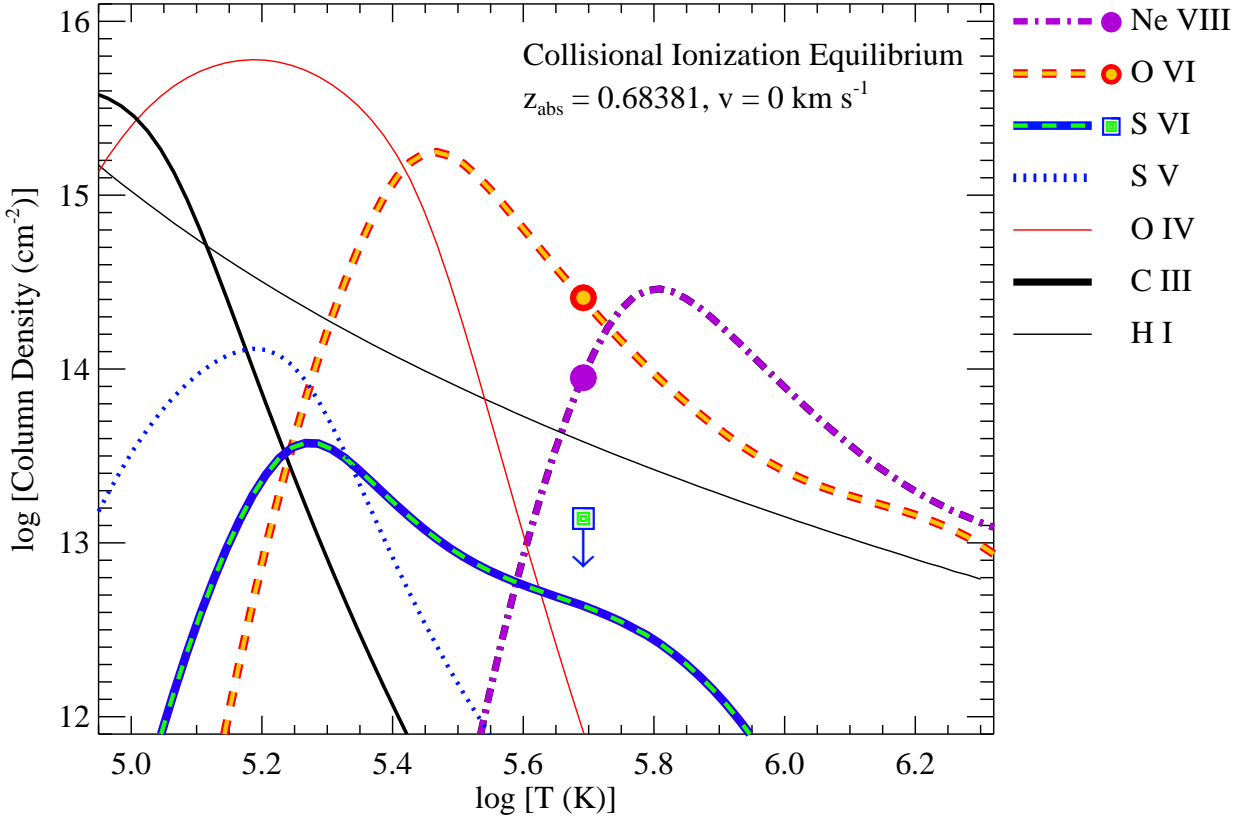}
\caption{Collisional ionization modeling of the highly ionized species
  detected in the $v = 0$ km s$^{-1}$ component of the PG1148+549
  absorber at $z_{\rm abs}$ = 0.68381, based on the collisional
  ionization equilibrium ion fractions calculated by
  \citet{Gnat07}. As in Figure~\ref{Fig:Ionization}, the column
  densities predicted by the model are shown with smooth curves, and
  the observed column densities of O~VI and Ne~VIII, as well as the
  observed $3\sigma$ upper limit on S~VI, are plotted with large
  symbols at T = $10^{5.7}$ K. The legend at right indicates the
  species that correspond to the curves and symbols.
  \label{Fig:Collion}}
\end{figure*}

\subsection{Collisional Ionization} 
\label{Sec:Collion}

\begin{table*}
\tablewidth{0pc}
\begin{center}
\caption{Properties of the Collisionally Ionized Hot Phase of the PG1148+549 Ne~VIII Absorbers \label{Tab:Hot}}

\begin{tabular}{lcccc}
\hline\hline
\vspace{0.1cm}
Redshift \ \ \  & log [T (K)] & log [N(H$_{tot}$) (cm$^{-2}$)] &  log [N(H I) (cm$^{-2}$)]$^{\rm a}$ & b (km s$^{-1}$)$^{\rm b}$ \\
\hline
0.68381\dotfill & 5.69        &   19.8                         &   13.6                              & 90                           \\
0.70152\dotfill & 5.69        &   $<19.0$                      &   $<12.8$                           & 90                              \\
0.72478\dotfill & 5.72        &   $<18.9$                      &   $<12.6$                           & 93                            \\
\hline
\end{tabular}
\end{center}
$^{\rm a}$ The predicted neutral hydrogen column density given the inferred temperature and N(H$_{tot}$) [see text]. \\
$^{\rm b}$ The predicted Doppler parameter for hydrogen at the given temperature. \\
\end{table*}

The photoionization models fail to produce enough
O~VI and Ne~VIII to match the observed column densities because the
required gas densities are too low and the cloud sizes are too large.
Consequently, we now consider whether addition of a hot, collisionally
ionized phase provides a viable explanation of the O~VI and Ne~VIII.  To predict column densities of various ions as a function
of temperature, we employ the ion fractions in collisionally ionized
gas from \citet{Gnat07}; we mainly assume that the gas is in
collisional ionization equilibrium (CIE), but we provide some brief
comments on how the results can change if the absorption arises in
non-equilibrium, rapidly cooling gas. We also must make assumptions
about the metallicity and total H (H~I + H~II) column density of the
absorbing gas [$N$(H$_{tot}$)].  The profiles of O~III, O~IV, O~VI,
and Ne~VIII are all well aligned in velocity space and there is
clearly some type of relationship between the Ne~VIII phase and the
lower ionization material, so we assume that the hot gas has the same
metallicity as the photoionized gas.  With that assumption, we find
the gas temperature that matches the Ne~VIII/O~VI ratio and adjust
$N$(H$_{tot}$) to fit the observed columns of the metals.
Table~\ref{Tab:Hot} lists the results of this procedure (in the CIE
case) for the $v = 0$ km s$^{-1}$ components of the absorbers at
$z_{\rm abs}$ = 0.68381, 0.70152, and 0.72478, including the derived
temperature and $N$(H$_{tot}$) as well as the predicted column density
of H~I in the hot gas and the H~I Doppler parameter.

As an example, Figure~\ref{Fig:Collion} shows the predicted CIE column
densities of several relevant ionization stages in the $v = 0$ km
s$^{-1}$ component of the Ne~VIII absorber at $z_{\rm abs}$ = 0.68381.
These hot-gas models have several notable features.  First, all of the
observed column densities, as well as upper limits on undetected
species, are fully consistent with a combination of a hot-gas phase
(the source of O~VI and Ne~VIII) and a distinct photoionized (cool)
gas phase (the source of all other detected species). 
In the $v = 0$ km s$^{-1}$ component at
$z_{\rm abs}$ = 0.68381, the Ne~VIII/O~VI ratio requires a gas
temperature of $T = 10^{5.7}$ K.  At this temperature, the predicted S~V
and S~VI column densities from the hot gas are both below the observed
upper limits.  The O~IV and C~III columns from the hot gas are also
extremely low, so we can legitimately assume that the O~IV and C~III
absorption only arises in the cool, photoionized phase.

On the other hand, we see from Figure~\ref{Fig:Collion} that even
though the Ne~VIII-bearing gas is hot, it does contain an appreciable
amount of H~I (thin black line in Figure~\ref{Fig:Collion}): the
hot-gas model predicts log $N$(H~I) = 13.6 for this component at the
inferred temperature and assumed metallicity.  However, H~I lines
arising in the hot phase will also be extremely broad ($b =
\sqrt{2kT/m}$) as indicated in Table~\ref{Tab:Hot}, and this makes
detection of the broad H~I absorption challenging.  Our COS spectrum
does not cover the Ly$\alpha$ line at the redshifts of the Ne~VIII
absorbers, and the broad Ly$\beta$ line predicted by the model in
Figure~\ref{Fig:Collion} is too weak to be detected in our data.  
Future observations of the Ly$\alpha$ line could, in
principle, detect the predicted broad Ly$\alpha$ line and thereby
corroborate our hot-gas detection.  Moreover, this would provide
additional information -- the combined analysis of $b-$values of
low-mass and higher-mass elements constrains the non-thermal line
broadening (e.g., due to turbulence) as well as the temperature (see
\S 4.1.1 in Tripp et al. 2008). So, while we have a good estimate of
the gas temperature of the hot phase of the Ne~VIII
absorbers, it would still be worthwhile to observe the Ly$\alpha$ line
to obtain more insight.
Absorption systems showing evidence of broad Ly$\alpha$ plus narrower
O~VI and/or Ne~VIII have been identified previously
\citep{Tripp01,Richter04,Nara10,Sav11a,Sav11,Nara12}, and analysis of
well-aligned H~I and O~VI absorption lines indicates that non-thermal
broadening is important \citep{Tripp08}.

We do not yet have the ability to apply such an analysis to the
systems in this paper, but we can at least check if the Doppler
parameters that the metal lines provide are consistent with gas
temperatures implied by the Ne~VIII/O~VI ratios.  For log T$\approx$
5.7 K, as indicated by the Ne~VIII/O~VI ratios in these systems, the
Doppler parameters from thermal broadening are $b$(O~VI) = 22.7 km
s$^{-1}$ and $b$(Ne~VIII) = 20.6 km s$^{-1}$.  The observed Doppler
parameters for all of the Ne VIII lines in these systems are in
agreement with this constraint.  The best fit Doppler parameters for
the O VI \lala 1031.93, 1037.62 lines in the \za = 0.68381 and \za =
0.70152 systems are also consistent with the expected line width.  The
O~VI line width in the \za=0.72478 system (which suffers from
particularly noisy profiles) is slightly more discrepant with the
expected line width but is still consistent within 2$\sigma$.

Finally, comparing the total hydrogen column densities derived for the
hot gas vs. the photoionized, cool gas (compare Table~\ref{Tab:Photo}
and \ref{Tab:Hot}), we see that total hydrogen column densities are
generally more than an order of magnitude higher in the hot gas than
in the cool gas, and therefore the hot (Ne~VIII) phase likely contains
substantially more mass than the photoionized gas.  This is similar to
the preponderance of hot gas found by \citet{Tripp11} in a multiphase
Ne~VIII absorber affiliated with a post-starburst galaxy.  This
conclusion rests on the assumption that the metallicity of the hot
phase is roughly the same as the metallicity of the cool phase.  If
the hot phase has a higher metallicity, it will have a lower total
hydrogen column, but we note that even if the hot phase metallicity is
ten times higher than that of the cool phase, the hot gas will still
harbor a similar total hydrogen column density.  It has been suggested
that in general, the cool circumgalactic gas detected in QSO absorbers
is an important baryon reservoir \citep{Werk13}, and it appears that
there is just as much mass (or more) in the hot gas traced by Ne~VIII.

We have also considered the non-equilibrium collisional ionization
models of \citet{Gnat07}. Again, with these models we find that no
single-phase model can account for the observed ratios and columns of
O~III, O~IV, O~VI, and Ne~VIII -- the non-equilibrium models also
require a distinct, hot phase.  In general, the non-equlibrium models
yield pretty similar results for the hot phase but can fit the
observed column densities at a slightly lower temperature (see, e.g.,
Figure S5 in Tripp et al. 2011).  It is possible that the O~VI
$b-$value in the system at $z_{\rm abs}$ = 0.72478, which is somewhat
more narrow than expected based on the CIE model, favors a
non-equilibrium situation.  However, higher S/N measurement of the
line width is needed before this can be considered to be strong
evidence.

\section{Affiliated Galaxies}
\label{sec:galaxies} 

\begin{figure}
\includegraphics[width=\linewidth]{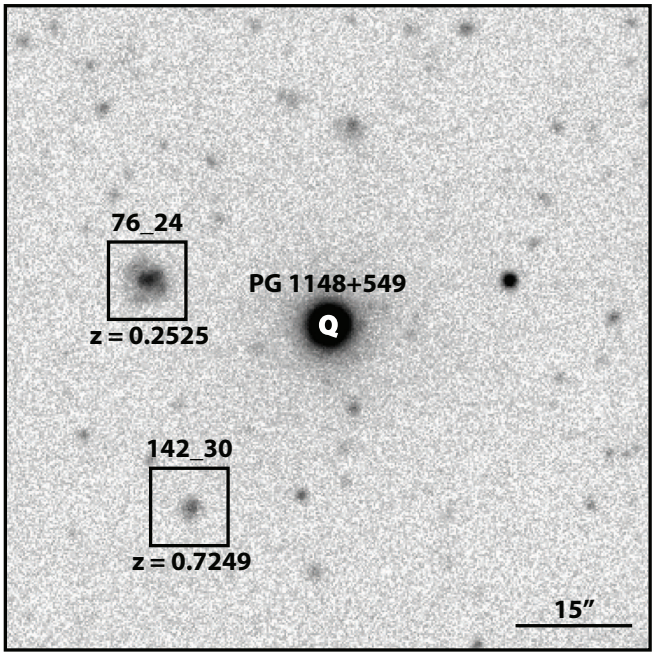}
\caption{Deep U band image of the PG1148+549 field from the Large
  Binocular Telescope.  The image is centered on the QSO. Galaxies for
  which we have obtained spectroscopic redshifts with Keck/LRIS are
  marked with a $10''$ box.  The label above each box is our galaxy
  identification code; the first number in the galaxy ID label is the
  position angle (north through east) from in the QSO (in degrees),
  and the second number is the angular separation from the QSO (in
  arcseconds). \label{Fig:Field}}
\end{figure}

\begin{figure*}
\includegraphics[width=\linewidth]{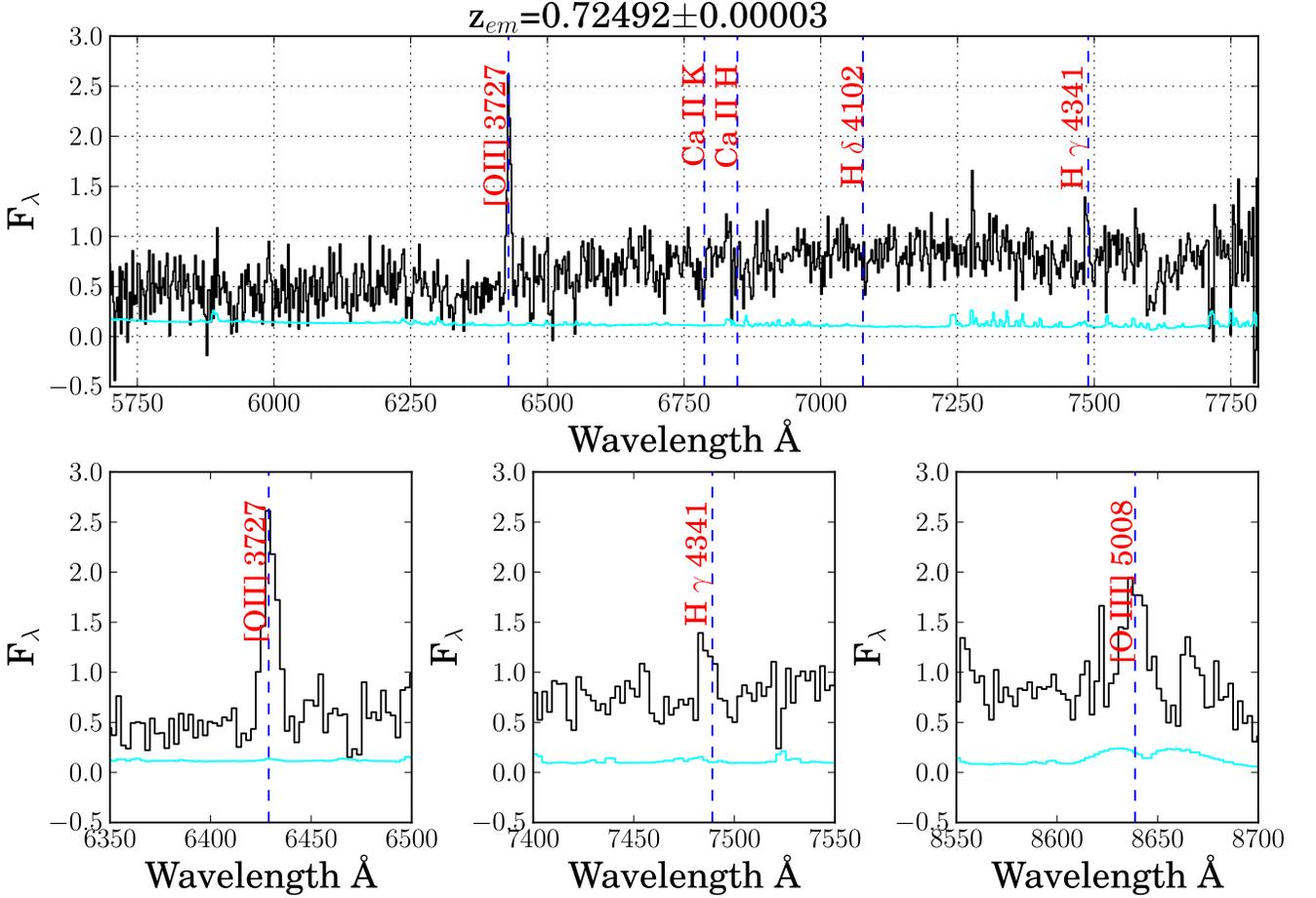}
\caption{Keck/LRIS spectrum of the galaxy 142\_30 (see
  Figure~\ref{Fig:Field}).  Several emission lines are evident
    including Balmer lines, [O~II], and [O~III], and these features
    indicate a vacuum Heliocentric redshift of $z_{em}=0.72492$.  The
    emission lines also provide constraints on the star-formation rate
    and metallicity of the galaxy (see Table~\ref{Tab:GalPars}).  The
    cyan line below the spectrum represents the 1$\sigma$ uncertainty
    in the flux.  \label{Fig:GalSpec}}
\end{figure*}

To begin to probe the relationships between the PG1148+549 Ne~VIII
absorbers and nearby galaxies, we have obtained deep imaging and
spectroscopy, as summarized in \S~\ref{Sec:Observations}.
Figure~\ref{Fig:Field} shows our $U$-band LBT image of the PG1148+549
field, which has a $5\sigma$ limiting magnitude of $U_{AB} = 26.0$. We obtained
Keck/LRIS spectra of the four brightest objects near the sight line.
The LRIS spectra showed that two
of the LRIS targets are galaxies at interesting redshifts.  These
galaxies are labelled in Figure~\ref{Fig:Field} by their position angle
from the QSO (North through East) and their angular separation from
the sight line.  As described in \citet{Werk12}, we have estimated the
star-formation rates, H~II-region metallicities, absolute magnitudes,
and impact parameters of the galaxies observed with LRIS.  These
properties are presented in Table~\ref{Tab:GalPars}, and the spectrum
of galaxy 142\_30 is shown in Figure~\ref{Fig:GalSpec}.

One of the LRIS galaxies (76$\_$24) is at a low enough redshift
($z_{\rm gal}$ = 0.2525) so that we cannot check for affiliated
Ne~VIII, but we note that the COS spectrum shows strong O~VI, C~III,
and multiple H~I Lyman series absorption lines at this redshift,
consistent with recent findings that strong O~VI absorption is
ubiquitious in the halos of star-forming galaxies
\citep{Pro11,Tum11b}.  The other LRIS galaxy (142\_30) is only $\Delta
v \sim+30$ \kms from the Ne~VIII absorber at \za=0.72478.  In many
regards, Galaxy 142\_30 has the characteristics one might expect to
find for an object embedded in a gas-rich halo: the galaxy is luminous
(L$\sim$L$^{\star}$), has a high metallicity ([O/H]=+0.22$\pm$0.15),
and a relatively high SFR of 6.42$\pm$0.44 M$_{\sun}$ yr$^{-1}$.
However, the galaxy has one surprising feature: it is fairly far from
the sight line (impact parameter = 217 kpc). Given the high
metallicity of the affiliated absorber, $Z > Z_{\odot}$, the
relatively large distance to the nearest galaxy is notable, although
we stress that more galaxy redshift measurements are required to
thoroughly probe the origin of this absorbing gas.  

\begin{table*}
\begin{center}
\tablewidth{0pc}
\caption{Spectroscopic Properties of Galaxies Near
  PG1148+549 \label{Tab:GalPars} }
\begin{tabular}{lccccc}
\hline\hline
Galaxy ID$^{\rm a}$   & $z_{\rm galaxy}$ & Impact Parameter   & Star-formation Rate$^{\rm b}$ & M$_{R}^{\rm c}$         & [O/H]$^{\rm b}$ \\
                  &                  & (kpc)              & (M$_{\sun}$ yr$^{-1}$)    &                     &       \\     
\hline
76$\_$24\dotfill  & 0.25250          & 95                 & 1.53$\pm$0.04             & $-19.86\pm0.04$     & $-0.11\pm0.15$ \\
142$\_$30\dotfill & 0.72492          & 217                & 6.42$\pm$0.44             & $-21.15\pm0.13$     & $+0.22\pm0.15$   \\
\hline
\end{tabular}
\end{center}

$^{\rm a}$ The first number in the galaxy identifier is the position
angle (north through east) in degrees, and the second number is the angular
separation from the QSO in arcseconds. \\
$^{\rm b}$ Star-formation rate and logarithmic metallicity measured as described in \citet{Werk12}. \\
$^{\rm c}$ $R$-band absolute magnitude. \\
\end{table*} 

\section{Discussion}
\label{sec:discussion}

\subsection{Redshift Density and Baryonic Content}

One of the primary goals of our COS survey is to provide some basic
statistics on Ne~VIII QSO absorption systems such as the number of
these systems per unit redshift, d$\mathcal{N}$/d$z$.  The Ne~VIII
absorption lines are weak, and at the redshifts where they can be
observed with \textit{HST}, there is a moderately high density of
absorption lines from other redshifts, so discovery of the Ne~VIII
absorbers is slow work that requires careful identification of all of
the lines in a spectrum (not just the Ne~VIII) to avoid spurious
misidentifications.  This work is underway, but we can obtain some
preliminary statistics based on the PG1148+549 data.

The COS spectrum of PG1148+549 enables detection of the redshifted
Ne~VIII \lala 770.41,780.32 lines above $z=0.49$. Here we are mainly
interested in the circumgalactic gas in the large halos of intervening
galaxies.  It is known that there is a statistically significant
excess of highly ionized absorbers with $z_{\rm abs} \approx z_{\rm
  QSO}$ \citep[see, e.g., Figures 14-15 in][]{Tripp08}, many of which
arise in AGN-driven outflows and other phenomena that occur quite
close to the central engine of active galactic nuclei
\citep[e.g.,][]{GangB08}.  For this reason, we excluded the region
within 3000 \kms of $z_{\rm QSO}$ from the part of the PG1148+549
spectrum that we probed for Ne~VIII absorption.  Following the
method outlined in \citet{Tripp08} to account for variable S/N and
regions blocked by strong lines from other redshifts, we find that the
total redshift over which we can detect Ne~VIII is $\Delta z=0.43$ for
a 10-pixel wide line and a minimum detection threshold of 30 m\AA . In
terms of the comoving ``absorption distance'' \citep{Bah69},
\begin{equation}
dX = \frac{H_{0}}{H(z)} (1+z)^2 dz,
\end{equation}
where $H_{0}$ is the Hubble constant and $H(z) = H_{0}\sqrt{\Omega
  _{m}(1+z)^{3} + \Omega _{\Lambda}}$ in our adopted cosmology, we can
detect Ne~VIII over a total comoving path of $\Delta X$ = 0.87.

With these numbers, we obtain d$\mathcal{N}$/d$z = 7^{+7}_{-4}$ for
Ne~VIII absorbers with W$_{0}>30$ m\AA\ or, in terms of the comoving
path, d$\mathcal{N}$/d$X = 3^{+4}_{-1}$, with uncertainties from
Poisson statistics \citep{Geh86}.  This estimate will be revised in
future work when all of the spectra from this program will be analzed
and presented. We note that the O~VI $\lambda$1031.93 lines affiliated
with these Ne~VIII absorbers have rest equivalent widths greater than
70 m\AA , and the d$\mathcal{N}$/d$z$ of O~VI absorbers with $W_{\rm
  r} > 70$ m\AA\ \citep{Tripp08,Tilton12} is similar to the
d$\mathcal{N}$/d$z$ that we find for Ne~VIII.

We can also estimate the cosmological mass density, $\Omega \equiv
\rho / \rho_{c}$, traced by the Ne VIII absorbers.  The cosmological
mass in the form of Ne~VIII ions only is
\begin{equation}
\Omega_{\rm Ne~VIII}=\frac{H_0}{c}\frac{m_{\rm Ne}}{\rho_c}\frac{\sum_{i}
  N_i({\rm Ne~VIII})}{\Delta X\ }
\end{equation}
where $m_{\rm Ne}$ is the mass of neon and $\rho_{c}$ is the critical density. Summing over the Ne~VIII absorbers discussed above,we obtain
$$
\Omega_{\rm Ne~VIII} \approx 6 \times 10^{-8}.
$$ 
The mass in Ne~VIII ions is tiny, but for some questions, this
quantity is a useful constraint.  Most of the mass in these absorbers
is, of course, in the ionized hydrogen, which can only be estimated
with models that constrain the hydrogen ionization correction.  For
this purpose, we use the hot-gas results presented in
Table~\ref{Tab:Hot}, and we calculate the total baryonic mass in the
Ne~VIII systems using
\begin{equation}
\Omega_{b}=\frac{H_0}{c}\frac{\mu
  m_{\rm H}}{\rho_c}\frac{\sum_{i} N_i({\rm H}_{tot})}{\Delta X\ },
\end{equation}
where $m_{\rm H}$ is the hydrogen mass and we assume $\mu$ = 1.3 to
account for mass in helium. From the total hydrogen column densities
in Table~\ref{Tab:Hot} and this expression we find
$$
\Omega_{b} \lesssim 0.002.
$$ This is an upper limit because two of the Ne~VIII systems provide
only upper limits on $N$(H~I) and $N$(H$_{tot}$).  It is interesting
to note that this upper limit amounts to $\approx 4$\% of the baryons
\citep{Sper07}, but we emphasize that both $\Omega _{\rm Ne~VIII}$ and
$\Omega_{b}$ are still quite uncertain due to the small sample and
uncertainty in the metallicity\footnote{The $N$(H$_{tot}$) derived
  from the models depends on the adopted metallicity of the
  Ne~VIII-bearing gas, which we have assumed is the same as the well
  constrained cooler, photoionized gas in each absorber.  If the
  Ne~VIII phase tends to be more mixed with lower metallicity ambient
  gas, then the metallicity of the Ne~VIII could be lower, which would
  increase $N$(H$_{tot}$) and $\Omega_{b}$. Conversely, if the hot gas
  (including the Ne~VIII) tends to separate from the cooler gas
  \citep[as occurs in some models, e.g.,][]{maclow99}, then the
  Ne~VIII gas could have higher metallicity.}  of the Ne~VIII-bearing
gas.  We will revisit these quantities in a future paper when our full
sample has been analyzed.

In addition to the large uncertainty due to the small sample, we note
one curious aspect of the PG1148+549 Ne~VIII absorbers: while our data
have sufficient signal-to-noise to detect these lines over a large
redshift range ($0.51 \leq z \leq 0.96$), all of the Ne~VIII systems
that we identify are clustered in a small portion of this redshift
window with $0.68381 \leq z_{\rm abs} \leq 0.72478$.  This raises a
question: could the three absorption systems be related somehow? For
example, could these absorbers all be ejecta from a single galaxy?
This seems unlikely because the absorber redshifts are spread over a
large velocity range (7200 km s$^{-1}$).  Outflows driven by quasars
(e.g., broad absorption-line outflows) do attain such velocities, but
usually these outflows exhibit many adjacent components spread over
the velocity range, or they show a smooth but very broad absorption
trough extending over a large velocity interval \citep[see,
  e.g.,][]{Capell12}.  The PG1148+549 Ne~VIII absorbers have no
resemblence to BAL outflows.  Some quasars have been shown to have
absorption systems due to QSO ejecta that are highly displaced in
redshift from the QSO \citep[e.g.,][]{JannuziXX, Bowen01, RodH11}, but
these also show distinctive characteristics (e.g., relatively broad
and smooth absorption profiles) that do not match the PG1148+549
NeVIII systems.

It seems improbable that these Ne~VIII absorbers originate in an
outflow from a single galaxy.  A more plausible explanation is that
the sight line passes through a region with a relatively high density
of gas-rich, star-forming galaxies at $z \approx$ 0.7, and each of the
three absorbers originate from separate galaxies.  This is a testable
hypothesis, and we are obtaining additional deep, multiobject
spectroscopy of a large number of galaxies in the PG1148+549 field
which could reveal such a large-scale structure.  This spectroscopic
survey is underway but not yet complete; once this has been completed,
we will reexamine the origins of these Ne~VIII systems.  
The existence of such a structure on this line of sight could
bias the redshift density and baryonic content estimates.
Incorporation of a larger number of sight lines is necessary to reduce
the effects of this type of cosmic variance, and analysis of these
statistics with larger numbers of sight lines is underway.
 
\subsection{Cross Section, Size, and Stability}

The redshift density (\dndz ) of the Ne~VIII absorbers can be
interpreted in a variety of ways.  Frequently, \dndz\ is used to
estimate an effective cross section ($\pi R^2$) of absorption systems:
\begin{equation}
\frac{d\mathcal{N}}{dz}=n_{\rm abs}\pi R^2 \frac{c}{H(z)}
(1+z)^2, \label{eqn:crosssec}
\end{equation}
where $n_{\rm abs}$ is the spatial (volume) density (number per Mpc)
of the absorbing entities (not the particle density).  While we have
an approximate determination of the product of $n_{\rm abs}$ and $\pi
R^{2}$, we do not know the values of these quantities individually.
Nevertheless, we have derived constraints on the the thickness of the
low-ionization component of the absorbers (final column of
Table~\ref{Tab:Photo}) -- the photoionization models imply that the
low-ionization phase has a thickness ranging from less than 100 pc up
to 3.5 kpc.  It is important to note that all of the intervening
Ne~VIII absorbers in the PG1148+549 sight line and all of the Ne~VIII
absorbers from other sight lines reported in the literature show well
detected lower ionization absorption lines that are well aligned with
the Ne~VIII.  Therefore, it appears that the cross section of the
Ne~VIII and the lower ionization stages are the same -- in order to
detect Ne~VIII, it has (so far) always been necessary for the sight
line to pierce the lower ionization cloud as well.  If we assume,
based on the sizes derived for the lower ionization phases, that the
clouds have a typical diameter of 1 kpc and we solve
equation~\ref{eqn:crosssec} for the space density, we find that
$n_{\rm abs} \approx 1100$ Mpc$^{-3}$.  This number is highly
uncertain; just considering the $1\sigma$ uncertainty in \dndz ,
$n_{\rm abs}$ could range from $\approx$ 500 up to 2000 Mpc$^{-3}$.
Nevertheless, even with these uncertainties, $n_{\rm abs}$ for these
absorbers is vastly larger than the space density of galaxies at $z
\approx$ 0.7.  Integrating the galaxy luminosity functions from the
DEEP2 or VVDS surveys \citep{Willmer06,Ilbert05} indicates that
galaxies with $L \gtrsim 0.1 L*$ have a space density of $n \approx
10^{-2}$ at this redshift.  Previous studies provide strong evidence
that O~VI absorbers are affiliated with galaxies
\citep{Stocke06,Pro11,Tum11b}, but their statistics require $> 10^{4}$
analogous clouds per galaxy to explain the observed \dndz .  This
requirement for a large number of clouds per galaxy has been noted for
other types of QSO absorbers including weak Mg~II systems at similar
redshifts \citep{Rigby02} and metal-rich C~IV absorbers at higher
redshifts \citep{Schaye07}.  An important caveat in this calculation
is that the density of the photoionized gas, and hence the absorber
thickness, depends on the assumed intensity of the photoionizing flux.
While models of the UV background and observational constraints on its
intensity appear to be in broad agreement \citep[see, e.g., Figure 7
  in][]{DT01}, it is possible that there is substantial spatial
variability of the UV background flux or systematic errors in its
estimation, and this could affect our estimate of $n_{\rm abs}$.
Indeed, detection of Ne~VIII may introduce a bias if these absorbers
are more likely to arise in regions of elevated flux.  However, even
with this uncertainty, it seems likely that any plausible UV ionizing
flux willl require $n_{\rm abs} \gg n_{\rm gal}$.

Alternatively, we can turn this around and ask: if the Ne~VIII
absorbers are affiliated with some type of galaxy of known spatial
density $n_{\rm gal}$, what is the effective cross section of the
Ne~VIII-bearing gas that produces the observed \dndz ? \citet{TumF05}
used this argument to show that if low$-z$ O~VI absorbers originate in
faint/dwarf galaxies, then metals must be transported to distances of
$\approx$200 kpc from their galaxies of origin (if these metals come
from $L*$ galaxies, they must be transported even farther).  As noted
above, galaxies with $L \gtrsim 0.1 L*$ have a spatial density of
$n_{\rm gal} \approx 10^{-2}$ Mpc$^{-3}$ at $z = 0.7$.  Equating this
value of $n_{\rm gal}$ to $n_{\rm abs}$ in
equation~\ref{eqn:crosssec}, we obtain $R \sim$ 100 - 200 kpc, where
the range reflects the uncertainty in \dndz .  Here we have assumed
that the covering fraction of the Ne~VIII systems is $\approx$ 1; if
the covering fraction is smaller, then the effective cross section
must be even larger.  While this is a rough initial estimate, this
implies a large transport distance for such metal-rich gas.  How is
solar-metallicity gas propagated to such a large distance?  We note
that the frequency of Ne~VIII absorbers may indicate that they are
more likely to be connected with lower luminosity dwarf galaxies,
based on an argument analogous to the one applied by \citet{TumF05} to
the O~VI systems: if the Ne~VIII absorbers originate in $L*$ galaxies,
then they must have very large effective cross sections that seem
improbable.  If we only consider galaxies with $L \gtrsim 1.0 L*$, for
example, then the galaxy density drops to $n_{\rm gal} \approx
10^{-3}$, which requires an effective cross section $\gtrsim 1$
Mpc. The finding of \citet{Mul09} and \citet{Chen09} that lower
redshift Ne~VIII absorbers are affiliated with sub-$L*$ galaxies
supports this implication of the statistics.

The arguments above suggest that the Ne~VIII systems arise in extended
gaseous halos that are filled with a large number of small
clouds. Would such a configuration be stable?  For the photoionized
portions of the absorbers, we can estimate the radial size of the
photoionized phase, assuming it is in hydrostatic equilibrium, given
the gas temperature, $N$(H~I), the fraction of the mass in gas, and
the H~I photoionization rate \citep[see equation 12
  in][]{Schaye01}. Assuming typical values for these parameters, we
find that the properties of the photoionized gas in
Table~\ref{Tab:Photo} imply that the radial size of the photoionized
material at $z_{\rm abs}$ = 0.68381 should be roughly $60 - 100$ kpc
if it is in hydrostatic equilibrium. For the systems at $z_{\rm abs}$
= 0.70152 and 0.72478, we find that similarly large clouds are
expected in hydrostatic equilibrium with sizes $\gtrsim$ 100 kpc. In
contrast, the photoionization models suggest that the absorbers are
vastly smaller (last column in Table~\ref{Tab:Photo}).\footnote{The
  particle densities and absorber thicknesses in Table~\ref{Tab:Photo}
  were derived assuming the absorber is ionized by the diffuse UV
  background from quasars.  A local ionizing source, which would cause
  the ionizing photon density to be higher, would require higher
  particle densities and even smaller clouds, so this would only
  exacerbate the discrepancy.}  This might indicate that the
photoionized clouds in this system are {\it not} in hydrostatic
equilbrium but rather are in the process of expanding and/or
evaporating.

Alternatively, they could be pressure-confined by the surrounding
medium.  The \textsc{cloudy} models indicate that the pressure in the
cool, photoionized phase is $p/k \approx$ 10 cm$^{-3}$~K.  If the
Ne~VIII phase surrounds and pressure confines the cool gas, then
pressure balance requires a particle density of $\approx 10^{-5}$
cm$^{-3}$.  While this is a reasonble density for halo gas, it may
require sizes that are unreasonably large when combined with
$N$(H$_{tot}$) for the hot gas (see Table~\ref{Tab:Hot}), and for this
reason it is unlikely that the Ne~VIII phase pressure confines the
cool gas.  This is not surprising -- the close alignment of the
Ne~VIII phase with the cooler gas suggests that the Ne~VIII arises in
an interface between the cool gas and a hotter (unseen) ambient
medium; this hypothesis has been favored in analyses of other Ne~VIII
systems as well \citep{Mul09,Nar11,Tripp11}.  In this scenario, the
absorber is likely not a stable entity but rather is in the process of
dissipating into the halo in which it is embedded, e.g., outflow
ejecta that is expanding as it moves away from its source.

\section{Summary}

Using a high signal-to-noise spectrum of the bright QSO PG1148+549
($z_{\rm QSO}$ = 0.9754) obtained with the Cosmic Origins
Spectrograph, we have discovered three new Ne~VIII absorption-line
systems at redshifts ranging from $z_{\rm abs}$ = 0.68381 to 0.72478.
These absorbers exhibit a variety of absorption lines from other
metals including C~III, N~IV, O~III, O~IV, and O~VI. In
this work we have reported on the physical conditions and ionization
of these multiphase Ne~VIII/O~VI absorbers, and we have discussed the
implications of measurements regarding the nature and origin of the
absorbers.  In summary, our primary conclusions are the following:
\begin{enumerate}
\setlength{\itemsep}{0pt}
\item A single-phase model in which all of the ions are produced via
  photoionization or collisional ionization fails to reproduce the
  observed column density ratios; a multiphase absorption system is
  clearly required to explain the range of detected ions (from O~III
  up to Ne~VIII). Moreover, the O~VI and Ne~VIII cannot originate in
  photoionized plasma -- models in which the high ions are produced
  via photoionization predict unphysically large cloud sizes, with
  absorption lines that would be highly spread out by broadening in
  the Hubble flow.  We conclude that the O~VI and Ne~VIII lines arise
  in collisionally ionized gas with T$\sim10^{5.7}$ K, while the low
  ions are produced via photoionization.
\item The oxygen abundances of the systems, as determined from the low
  ions and photoionization modeling, are all [O/H]$>-0.5$, higher than
  the typically assumed metallicity of the CGM/IGM of
  [X/H]$\sim-1.0$. The \za=0.70152 and \za=0.72478 systems require
  supersolar metallicities.  These are abundances in the cool gas, and
  we assume for our analysis that the hot gas probed by the Ne~VIII
  has a similar metallicity.  The hot gas could have a different
  metallicity than the cool phase, e.g., it could be more metal rich
  because hot gas tends to separate from the cooler material in
  outflows \citep[e.g.,][]{maclow99}. On the other hand, the
  metallicity of the major baryon reservoirs in the CGM and IGM may
  have much lower metallicities and we are not yet detecting those
  reservoirs; detection of such low-metallicity material in metal
  absorption lines may require significantly higher signal-to-noise
  spectroscopy.
\item The absorbers also indicate moderate underabundances of carbon
  and nitrogen (compared to oxygen) at a level similar to the C and N
  underabundances observed in Milky Way halo stars.
\item Assuming that a portion of the gas is photoionized by the UV
  background from quasars and galaxies, the ionization models indicate
  that the low-ionization phase has low density ($\approx 10^{-4}$
  cm$^{-3}$), small size (3.5 kpc down to $< 100$ pc), and low thermal
  pressure, $p/k \approx$ 10 cm$^{-3}$ K.  The total (H~I + H~II)
  hydrogen column densities in these absorbers ranges from $<
  10^{17.1}$ cm$^{-2}$ up to $10^{18.5}$ cm$^{-2}$.
\item The hot, collisionally ionized phase that contains O~VI and
  Ne~VIII has T$\approx 10^{5.7}$ K based on the O VI to Ne VIII
  ratio, assuming collisional ionization equilibrium.  Non-equilibrium
  models \citep{Gnat07} require only slightly lower temperatures.  The
  ionization models indicate that the total hydrogen column density in
  the hot gas is approximately $10\times$ higher than the total
  hydrogen column in the cool, photoionized material.
\item The COS spectrum has sufficient S/N to reveal Ne~VIII lines with
  $W_{\rm r} > 30$ m\AA\ over a total redshift window of $\Delta z$ =
  0.43 (or a comoving absorption distance $\Delta X = 0.87$), which
  implies a redshift density of d$\mathcal{N}$/d$z = 7^{+7}_{-4}$
  (comoving redshift density d$\mathcal{N}$/d$X = 3^{+4}_{-1}$).  The
  cosmological mass density of the Ne~VIII ions is $\Omega _{\rm
    Ne~VIII} \approx 6 \times 10^{-8}$, and a preliminary constraint
  on the baryon mass in these absorbers is $\Omega _{b} \lesssim
  0.002$.
\item Given the small sizes implied for the low-ionization phase that
  is always detected along with the Ne~VIII, the observed
  d$\mathcal{N}$/d$z$ indicates that galaxies generally have a large
  number of these metal-enriched, multiphase clouds in their halos
  with effective cross sections of 100 kpc or more.
\item However, the clouds are unlikely to be stable and long-lived.
  The small size of the photoionized phase implies that the clouds
  will expand and dissipate, and the correlation of the Ne~VIII phase
  with the photoionized phase suggests that the Ne~VIII is the
  transitional layer produced as the cool gas is heated and
  photoionized, ultimately destined to blend into the hotter ambient
  halo of the circumgalactic medium.
\end{enumerate}

\section*{Acknowledgments}
We thank Hsiao-Wen Chen for comments on the manuscript, and we thank
Christina Williams and Marcel Neeleman for technical assistance. This
research was supported by NASA grants HST-GO-11741 and NNX08AJ44G.
J.X.P. also appreciates support from NSF grant AST-0709235. The
W. M. Keck Observatory is operated as a scientific partnership between
the California Institute of Technology, the University of California,
and the National Aeronautics and Space Administration. The Keck
Observatory was made possible by the generous financial support of the
W. M. Keck Foundation. The authors wish to recognize and acknowledge
the very significant cultural role and reverence that the summit of
Mauna Kea has always had within the indigenous Hawaiian community. We
are most fortunate to have the opportunity to conduct observations
from this mountain.  This work also made use of data from the Large
Binocular Telescope. The LBT is an international collaboration among
institutions in the United States, Italy, and Germany. The LBT
Corporation partners are: The University of Arizona on behalf of the
Arizona university system; Istituto Nazionale di Astrofisica, Italy;
LBT Beteiligungsgesellschaft, Germany, representing the Max Planck
Society, the Astrophysical Institute Potsdam, and Heidelberg
University; The Ohio State University; The Research Corporation, on
behalf of The University of Notre Dame, University of Minnesota, and
University of Virginia.

\end{document}